\documentclass[twocolumn,preprintnumbers,superscriptaddress,aps,nofootinbib]{revtex4-2}

\usepackage{amsmath,amssymb,mathtools}
\usepackage{graphicx}
\usepackage{booktabs,cellspace}
\usepackage{color}
\usepackage{hyperref}
\usepackage{xspace}
\usepackage{bm}
\usepackage[normalem]{ulem}
\usepackage{soul}
\usepackage{mathrsfs}
\makeatletter 
\renewcommand\onecolumngrid{% <<<<<<
\do@columngrid{one}{\@ne}%
\def\set@footnotewidth{\onecolumngrid}% <<<<<<<<<<<<<<<<
\def\footnoterule{\kern-6pt\hrule width 1.5in\kern6pt}%
}
\makeatother

\newcommand{\IPhTAff}{Universit\'{e} Paris-Saclay, CNRS, CEA, Institut de physique th\'{e}orique, 91191, Gif-sur-Yvette,
France}
\newcommand{\ISTAff}{Instituto Superior Técnico (IST), Universidade de Lisboa,
Av. Rovisco Pais 1, P-1049-001 Lisboa, Portugal}
\newcommand{\LIPAff}{LIP, Av. Prof. Gama Pinto, 2, P-1649-003 Lisboa, Portugal}
\newcommand{\GranadaAff}{Departamento de Física Teórica y del Cosmos, Universidad de Granada, Campus de Fuentenueva, E-18071 Granada, Spain}

\definecolor{darkgreen}{rgb}{0,0.5,0}
\definecolor{darkblue}{rgb}{0,0,0.7}
\definecolor{darkred}{rgb}{0.5,0,0.0}
\definecolor{darkorange}{rgb}{0.8,0.4,0.0}

\newcommand{\bp}{\bm{p}}

\def\cK{{\cal K}}
\def\cQ{{\cal Q}}

\def\Re{\text{Re}}

\begin{document}

\title{In-medium QCD splittings beyond the soft, large-$N_c$ and harmonic-oscillator approximations all at once}

\author{Marco Leitão} \affiliation{\IPhTAff} \affiliation{\ISTAff} \affiliation{\LIPAff}
\author{José Guilherme Milhano}\affiliation{\ISTAff}\affiliation{\LIPAff}
\author{Alba Soto-Ontoso}\affiliation{\GranadaAff}

\begin{abstract}
Nearly thirty years ago, Baier, Dokshitzer, Mueller, Peigné, Schiff, and Zakharov (BDMPS-Z) introduced a formalism to calculate the fully differential probability for a high-energy quark or gluon to radiate inside a finite-volume QCD plasma.    
We report on the first complete numerical solution of the BDMPS-Z equations for in-medium QCD splittings. Our numerical routines are precise across phase-space, enabling a determination of the in-medium splitting functions that is significantly beyond the state-of-the-art, including finite-energy effects, subleading-color contributions, and a realistic model for parton-medium interactions. 
We quantify the uncertainties associated with standard approximations in the literature, revealing substantial deviations across phase-space.
This work opens a path toward more precise calculations of jet observables and for powerful new constraints of medium parameters from high-energy heavy-ion collider data.
\end{abstract}

\maketitle
Extracting the material properties of quark-gluon matter at high temperatures is a long-standing objective of the heavy-ion collider programme. Among the most powerful probes of these properties are highly energetic particle jets, which are born from hard scatterings and evolve over space-time scales spanning several orders of magnitude before reaching the detector~\cite{Apolinario:2022vzg,Wang:2025lct}. 
This multi-scale sensitivity has been instrumental in discovering and characterizing the deconfined, chirally symmetric phase of matter known as the quark-gluon plasma (QGP)~\cite{RevModPhys.90.025005,Cunqueiro:2021wls,ALICE:2022wpn,CMS:2024krd}. 
The precision frontier enabled by the high-luminosity LHC era offers unprecedented opportunities to leverage jet observables for decoding the microscopic structure of hot, quark-gluon matter~\cite{Citron:2018lsq}. 

\begin{figure}[t] \centerline{
\includegraphics[width=0.95\columnwidth]{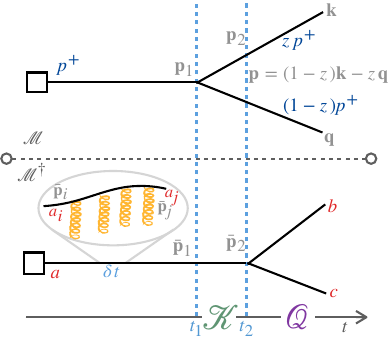}}
 \caption{Diagrammatic representation of the time evolution of the splitting kernel computed in this Letter, as described by Eq.~\eqref{eq:full-spectrum}. The dashed gray line separates amplitude from its complex conjugate. The inset illustrates how transverse momentum and color evolve at each time step due to interactions with the medium quasi-particles, while the light-cone longitudinal momentum $p^+$ is conserved.}
 \label{fig:sketch}
\end{figure}

The theoretical description of jet evolution in the QGP is a multidisciplinary endeavor that integrates insights from both perturbative and non-perturbative techniques in the context of Quantum Chromodynamics (QCD)~\cite{Mehtar-Tani:2025rty}. At the heart of this framework is the evaluation of in-medium QCD splitting kernels, which encode the probability for a high-energy parton, $a$, to branch into two daughter partons, $b$ and $c$, within the medium, as illustrated in Fig.~\ref{fig:sketch}. These splitting kernels serve as building blocks for calculating jet observables, whether through Monte-Carlo event generators or via analytical calculations. This work focuses on developing a method to compute these splitting kernels that is simultaneously precise and numerically efficient across the entire phase space.

The foundational BDMPS-Z framework~\cite{Baier:1994bd,Baier:1996kr,Baier:1996sk,Zakharov:1996fv,Zakharov:1997uu} established a path-integral formulation for in-medium QCD splitting kernels under several simplifying assumptions. At the kinematic level, longitudinal momenta are assumed to be much larger than any characteristic transverse scale (\textit{close-to-eikonal approximation}). Concurrently, the QGP is modeled as a finite-volume, stochastic color field generated by static quasiparticle sources. Despite these simplifications, the sheer complexity of the resulting equations has so far restricted their solution to specific limiting cases. One class of approximations constrains the splitting kinematics, such as the \textit{soft} (\textit{hard}) limit, where the energy fraction $z$ of the emitted parton is much smaller than (very close to) that of the parent parton, i.e. $z\to 0$ ($z\to 1$)~\cite{Blaizot:2012fh,Apolinario:2012vy}. Another class of approximations simplifies the color structure of the splitting by taking the number of colors in QCD to be large (\textit{large-$N_c$ approximation}) so as to reduce the complexity of the color algebra. A third class of approximations involves the modeling of parton-medium interactions; earlier calculations typically assumed either multiple soft momentum exchanges (\textit{harmonic-oscillator approximation})~\cite{Gyulassy:1993hr,Gyulassy:2002yv,Arnold:2009mr,Armesto:2003jh} or a single hard momentum exchange (\textit{$N=1$ opacity expansion})~\cite{Wiedemann:2000za,Gyulassy:2000fs,Djordjevic:2003zk,Ovanesyan:2011kn}, or an infinite QGP~\cite{Arnold:2002ja,Jeon:2003gi}. More recently, substantial progress has been made towards relaxing these assumptions by incorporating finite-$z$ corrections~\cite{Apolinario:2014csa,Dominguez:2019ges,Isaksen:2020npj,Sievert:2018imd,Andres:2026qrt}, subleading-$N_c$ contributions~\cite{Isaksen:2020npj,Barata:2026icn}, and more realistic parton-medium interaction models~\cite{Feal:2018sml,Feal:2019xfl,Mehtar-Tani:2019tvy,Andres:2020vxs,Barata:2021wuf,Isaksen:2022pkj,Barata:2023qds,Kuzmin:2023hko,Kuzmin:2024smy}. However, these efforts have been strictly limited to lifting one (or at most two) approximation(s) at a time, leaving the problem of solving the BDMPS-Z equations in their full generality completely open.

This Letter introduces a novel numerical approach to solve the BDMPS-Z equations for arbitrary $z$ at finite $N_c$, fully resumming jet-medium interactions to all orders in opacity with a realistic scattering potential. The starting point of our formulation is the established expression for the differential emission spectrum of the process depicted in Fig.~\ref{fig:sketch} 
\begin{align}
\label{eq:full-spectrum}
\nonumber & \frac{dI}{dzd^2\bm p} = \frac{\alpha_sP_{ba}(z)}{4\pi^2\omega^2}\Re\Big[\int_{\bm p_1, \bm p_2,\bm{\bar p}_2, \bar{\bm{p}}} 
 \int_0^\infty d t_1  \int_{t_1}^\infty d t_2 \, \delta^{(2)}_{\bar{\bm{p}}+\bm{p}}
\\
&\times(\bm{p}_1\cdot \bar{\bp}_2)\, \cK(\bm{p_2},\bm{p_1}|t_2,t_1)
\cQ(\bm p,\bar{\bm{p}},\bm{p_2}, -\bar{\bm{p}}_2|t_\infty, t_2)\Big] \, ,
\end{align}
where $\bm p$ is the relative transverse momentum, $\int_{\bm{p}} \equiv \frac{d^2\bm{p}}{(2\pi)^2}$, $p^+$ is the light-cone longitudinal momentum of the parent parton, $\omega=z(1-z)p^+$, and $P_{ba}(z)$ is the vacuum Altarelli-Parisi splitting function. The strong coupling constant is denoted by $\alpha_s$, which we approximate as constant, and the intermediate momentum variables are defined as shown in Fig.~\ref{fig:sketch}. The objects $\cK(t_2,t_1)$ and $\cQ(t_\infty, t_2)$ can be obtained by solving either: (i) a path integral over medium-averaged traces of Wilson lines~\cite{Zakharov:1996fv}, (ii) Dyson-like integral equations~\cite{Caron-Huot:2010qjx}, or equivalently, (iii) a system of non-Hermitian Schrödinger-like equations for the corresponding Green's functions~\cite{Kovner:2003zj}. We opt for the third strategy, which requires a series of manipulations to Eq.~\eqref{eq:full-spectrum}.

First, considering a medium of finite length, $L$, we can decompose the cross section into a sum over three distinct time intervals, depending on the ordering between the medium length and the light-cone times for the splitting in amplitude ($t_1$) and complex-conjugate amplitude ($t_2$). Naturally, when $t_1, t_2>L$ we recover the vacuum splitting kernel, meaning that Eq.~\eqref{eq:full-spectrum} reduces to    
\begin{align}
\label{eq:vac}
\frac{dI^{\rm vac}}{dzd^2\bm p} \equiv \frac{dI}{dzd^2\bm p}\Theta(t_1>L)\Theta(t_2>L)=\frac{\alpha_sP_{ba}(z)}{2\pi^2\bm p^2} \, .
\end{align}
For the remaining contributions, we first compute their ratio with respect to Eq.~\eqref{eq:vac} to isolate the medium modifications, denoting these quantities by ${\cal R}(z,\bm p)$ such that the spectrum reads 
\begin{align}
\label{eq:spectrum-split}
 \frac{dI}{dzd^2\bm p} = \,& \frac{dI^{\rm vac}}{dzd^2\bm p}\,\Big[1+{\cal R}^{\text {in-in}}(z,\bm p)+{\cal R}^{\text {in-out}}(z,\bm p)\Big]\,\nonumber \\ 
 \equiv \,&   \frac{dI^{\rm vac}}{dzd^2\bm p} \, {\cal R}_{\mathrm{med}}(z, \bm{p}).
\end{align}
The two contributions to ${\cal R}_{\mathrm{med}}$ are calculated as follows. In the $t_1<L<t_2$ region Eq.~\eqref{eq:full-spectrum} reduces to
\begin{align}
\label{eq:in-out}
\nonumber {\cal R}^\text{in-out}(z,\bm p) &= -\frac{1}{\omega} \mathrm{Re} \Bigg[\int_0^Ldt_1\int_{\bm p_1}i\,(\bm{p}\cdot\bm{p}_1) \, \mathcal{K} (\bp, \bp_1|L, t_1) \Bigg] \\
& \equiv \frac{1}{\omega} \Re[\bm p \cdot {\cal A} (t=L,\bm p)]\, ,
\end{align}
where ${\cal A}$ is obtained by solving the following evolution equation in 2+1 dimensions
\begin{equation}
    i\partial_t  {\cal A}(t, \, \bm{{p^\prime}}) = \Bigg[\dfrac{{\bm{p^\prime}}^2}{2 \omega} - i(\tilde{v}_{ba}  \ast \cdot )\Bigg]{\cal A}(t, \bm{p^\prime}) -\bm{p^\prime}\, .
    \label{eq:schro-A}
\end{equation}
The operator $(\tilde{v}_{ba}  \ast \cdot)$ denotes a convolution in momentum space between the potential and the function $\cal A$, where the potential is process-dependent and encodes the elastic scattering cross section within the medium. Our algorithm is agnostic to the specific scattering potential model. The results in the main text adopt the Gyulassy-Wang (GW) description of the medium as a collection of static scattering centers~\cite{Wang:1992qdg}. In the soft limit, this potential reduces to a simple form in coordinate space~\cite{Barata:2020rdn}
\begin{align}
\label{eq:vba-coord-space}
v_b({\bm x}) &= \int_{\bm p} e^{i {\bm p}\cdot {\bm x}}\,\lim_{z\to 0}\tilde v_{ba}(z,\bm p) \, \nonumber\\
 &= \dfrac{1}{4}\tilde{q}\,C_{b}\, \bm{x}^2\left[\ln\dfrac{Q_s^2}{\mu_\star} +\ln\dfrac{1}{{\bm x}^2Q_s^2}\right ] + \mathcal{O}(\mu_\star^2\bm{x}^4) \, ,
\end{align}
where $\mu_\star$ is an infrared scale reflecting the screening nature of the medium and $C_{b}$ the Casimir factor corresponding to the soft prong $b$. The saturation scale, $Q_s$, represents the average transverse momentum acquired by the propagating probe via medium-interactions during its formation time. This scale is determined by solving the transcendental equation $Q_s^2 = \tilde q C_b L \ln Q_s^2/\mu^2_\star$~\cite{Barata:2020rdn}. Finally, $\tilde q$ is the bare jet-quenching transport parameter normalized to the corresponding color factor; for instance, $\tilde q= 4\pi\alpha_s^2 n$ within the GW model of the medium with constant density $n$. 

The $t_{1,2}<L$ contribution to the spectrum reads
\begin{equation}
\label{eq:in-in}
{\cal R}^\text{in-in}(z,\bm p) \equiv \frac{\bm p^2}{\omega^2} \, \Re \, \left[{\cal B} (t=L,\bm p, \bm{\bar p})\right|_{\bm{\bar p} = -\bm p}] \, ,
\end{equation}
where the definition of $\cal{B}$ can be directly inferred from Eq.~\eqref{eq:full-spectrum} by restricting the time integrals to the medium length. 
The evolution of $\cal{B}$ is dictated by a set of coupled differential equations, as the system oscillates between all possible color states during its time evolution. We label the different color configurations as $\cal{B}_\sigma$, using the convention $\sigma = 1$ for the state of physical interest. In practice, $\sigma \in \{ 1, 2\}$ for $g\to q\bar q$, $\sigma \in \{ 1, 2, 3\}$ for $q\to gq$ and $\sigma \in \{1,...,6\}$ for $g\to gg$. The resulting equation for $\cal{B}_\sigma$ is (4+1)-dimensional and takes the form
\begin{align}
    i\partial_t  {\cal B}_\sigma(t, \, \bm{k},\,\bm{l}) &=    \left[
        \,\frac{2\bm{k}\!\cdot\!\bm{l}}{\omega}\,
        \delta_{\sigma}^{\sigma'}
        + i\left(\tilde{\mathbb{M}}_{\sigma}^{\sigma'}  \ast \cdot \right)
    \right]
    \mathcal{B}_{\sigma'}\!\left(t, \bm{k},\,\bm{l}\right)\nonumber \\
    &+
    (\bm{k} - \bm{l}) \cdot 
    {\cal A}\!\left(\bm{k}+\bm{l}\right)
    \, \xi_{\sigma} \, ,
    \label{eq:schro-B}
\end{align}
where $\tilde{\mathbb{M}}_{\sigma}^{\sigma'}$ is an evolution matrix, which contains process-specific information on the strength of the color mixing and diagonalizes by blocks in the large-$N_c$ limit. Another ingredient entering the previous system of equations is $\xi_{\sigma}$, a vector whose elements are unity when the associated correlator of Wilson lines has a non-zero initial value and zero otherwise. Since the solution lives on the anti-diagonal $\bar{\bm{p}} = -\bm{p}$, we switch variables to $\bm{k}= (\bm{p} - \bm {\bar p})/2$ and $\bm{l} =(\bm{p}+\bm {\bar p})/2$. This choice grants more flexibility in tuning the grid settings along the $\bm{l}$ axis, since only the subspace $\bm{l} = \bm{0}$ carries the physics. Furthermore, as discussed in the Supplemental Material, the rotational symmetry of the medium can be exploited in these new variables to reduce the dimensionality of the differential equations for both $\mathcal{A}$ and $\mathcal{B}$. This allows us to write ${\cal A}(t,p=|\bm p|)$ and ${\cal B}(t,k=|\bm k|,l=|\bm l|, \psi)$, where $\psi$ is the angle between $\bm k$ and $\bm l$. 

The evolution equations for $\cal{A}$ and $\cal{B}$ can be recast into the compact form 
\begin{equation}
    i\partial_t\bm{f} = \hat{\mathcal{H}}\bm{f} + \bm{s},\label{eq:evolution}
\end{equation}
where $\hat{\mathcal{H}} = \hat{\mathcal{T}} + i\hat{\mathcal{V}}$ is an effective, non-Hermitian Hamiltonian and $\bm{s}$ represents a source term. 
For a time-independent Hamiltonian, the formal recursive solution of Eq. \eqref{eq:evolution} reads
\begin{equation}
\label{eq:f-evol}
    \bm{f}(t+\delta t) = e^{-i\hat{\mathcal{H}}\delta t}\bm{f}(t) \, -i \, \int_{t}^{t+\delta t} d t' \, e^{-i\hat{\mathcal{H}}(t'-t)} \, \bm{s}(t'),
\end{equation}
where the $t'$-integral is evaluated using a quadrature scheme. Both contributions to Eq.~\eqref{eq:f-evol} require evaluating terms of the form $e^{-i\mathcal{H}\delta \tilde{t}}\bm{f}$ for various $\delta \tilde{t}$, which constitutes the central computational task of this work. 
Explicit numerical schemes, such as Runge-Kutta (RK) methods, approximate this exponential operator via a truncated polynomial expansion in powers of $\delta \tilde{t}$.
Consequently, the accuracy of such methods is strictly tied to the step size, requiring a stringent stability condition to prevent exponential error propagation.  For the equations at hand, this condition is neither available in closed form nor easily guaranteed, rendering explicit methods difficult to control over long propagation times, as previously noted in Ref.~\cite{Isaksen:2023nlr}.

To overcome these limitations, we opt for a spectral method based on a Faber polynomial expansion of the evolution operator, building on techniques from electrodynamics~\cite{BORISOV2006391} and condensed matter physics~\cite{Diogo_Soares_2024}. This approach scales the Hamiltonian, $\mathcal{H}_{\rm{sc}} \equiv \mathcal{\hat{H}}/\lambda$, ensuring its spectrum lies within a complex ellipse of unit logarithmic capacity. The Faber expansion takes the form 
\begin{equation}
    e^{-i\mathcal{H}\delta \tilde{t}}\bm{f} = \sum_{n=0}^{\infty}c_n(\delta \tau) \,\Phi_n(\mathcal{H}_{\rm sc}) \bm{f}, 
\end{equation}
where $\delta \tau \equiv \lambda \delta \tilde{t}$.
Both the scaling factor $\lambda$ and the expansion coefficients $c_n$ are determined using the spectral bounds of $\hat{\mathcal{H}}$, which can be inferred either from the structural properties of the operator or via numerical diagonalization techniques. The action $\Phi_n(\mathcal{H}_{\rm sc}) \bm{f}$ is evaluated using the recurrence relation of Faber polynomials tailored to the bounding ellipse. This series converges geometrically, with the truncation error controlled by the magnitude of the coefficients: truncating at the smallest $n$ for which $|c_{n}| \lesssim  \varepsilon$ bounds the relative error of approximating $e^{-i\mathcal{H}\delta \tilde{t}}\bm{f} $ to $\mathcal{O}(\varepsilon)$. In practice, $n \sim \mathcal{O}(10)$ suffices for moderate step sizes that are substantially larger than those typically needed by RK methods for a comparable accuracy. This approach decouples the expansion accuracy from the step size and, crucially, removes the step-size stability constraint, thereby enabling a robust and stable solution for $\cal{A}$ and $\cal{B}$ across the entire kinematic phase-space and range of medium parameters.

\begin{figure}[t] \centerline{
\includegraphics[width=0.98\columnwidth]{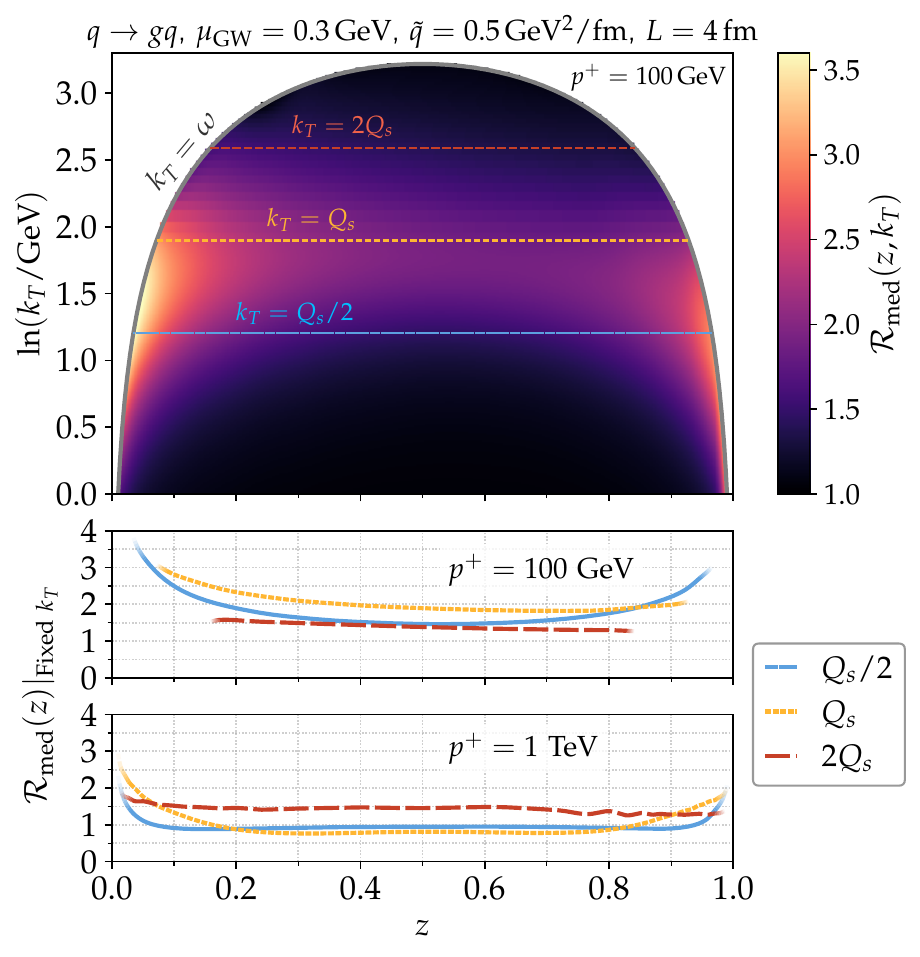}}
 \caption{Results for the medium modification factor $\mathcal{R}_{\mathrm{med}}(z,k_T=|\bm{p}|)$ in $q\to gq$ splittings. The middle (lower) panel shows $\mathcal{R}_{\mathrm{med}}(z)$ for $k_T =\lbrace Q_s/2, Q_s, 2Q_s \rbrace$ and $p^+ = 100 \, \mathrm{GeV}$ ($p^+ =1 \, \mathrm{TeV}$).}
 \label{fig:dome}
\end{figure}

With the above elements in place, we are now in a position to numerically evaluate the medium modification factor, ${\cal{R}}_{\mathrm{med}}(z,k_T=|\bm{p}|)$, for the splitting cross-section. The Supplemental Material contains a series of validation plots where we benchmark our numerical routines against semi-analytic results in specific limits, finding excellent agreement. Here, we focus on results that go beyond standard approximations intrinsic to previous approaches. For instance, in what follows we employ the GW model for the medium description and incorporate finite-$N_c$ corrections, using $C_A=N_c=3$ and $C_F=(N_c^2-1)/(2N_c)$. We adopt an LHC-guided selection of kinematic and medium parameters: $p^+=0.1, 1$ TeV to mimic the experimental coverage of the ALICE and ATLAS/CMS detectors, respectively, and choose the medium properties to roughly correspond to a central heavy-ion collision. 

Figure \ref{fig:dome} shows a Lund-style representation~\cite{Andersson:1988gp} of ${\cal{R}_{\mathrm{med}}}(z,k_T=|\bm{p}|)$ for $q\to gq$ splittings. A result of unity (${\cal{R}_{\mathrm{med}}}=1$) indicates that the medium does not modify the splitting process. This occurs in two specific regimes: either when $k_T\gg Q_s$ (where $Q_s\sim 7$ GeV for this parameter set), or when $k_T\ll Q_s$ at finite-$z$. In the bulk of phase-space, medium corrections exhibit a non-trivial kinematic dependence. These corrections increase significantly for unbalanced splittings, reaching a factor of $3\,(3.5)$ when $z\to 1\,(0)$. Conversely, as $p^+$ increases, ${\cal{R}_{\mathrm{med}}}$ decreases and can even fall below unity for $k_T \sim Q_s$, which represents a medium-induced suppression of the scattering rate. Similar qualitative behavior is found for $g\to gg$ and $g \to q\bar{q}$ splittings except for the $z \leftrightarrow 1-z$ asymmetry, as discussed in the Supplemental Material. Our results reveal that ${\cal{R}_{\mathrm{med}}}$ exhibits a complex parametric dependence on both medium properties and splitting kinematics, making it challenging to model with semi-analytic approaches. 

\begin{figure}[t] \centerline{
\includegraphics[width=0.99\columnwidth]{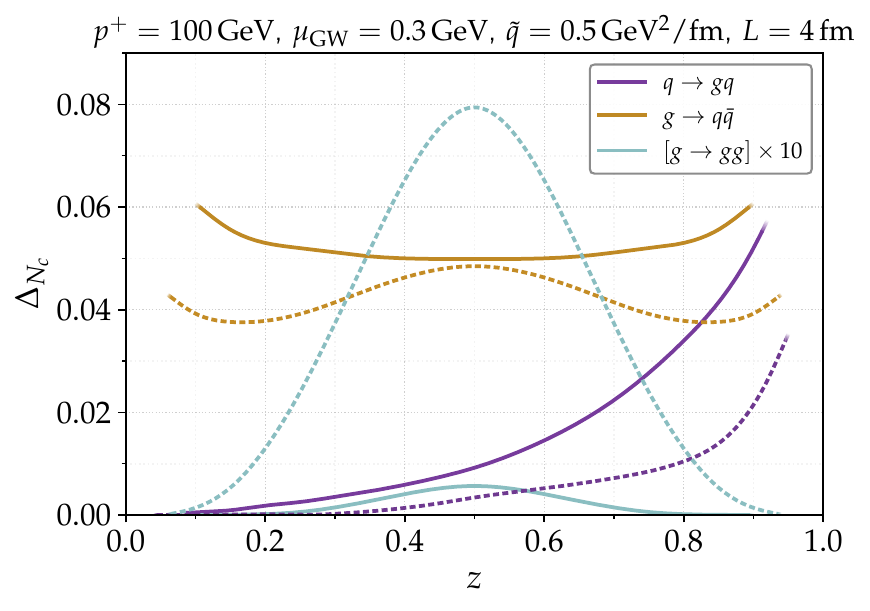}}
 \caption{Summary of $N_c$-corrections across splitting channels as a function of $z$ for $k_T = Q_s/2$ (dotted lines) and $k_T = Q_s$ (solid lines). For readability purposes we multiply the $g\to gg$ results by a factor $10$.}
 \label{fig:nc}
\end{figure}

Finite-$N_c$ corrections as a function of $z$ at $k_T = \lbrace Q_s/2, Q_s\rbrace$ for all splitting channels are shown in Fig.~\ref{fig:nc}. To quantify these corrections, we define the metric 
\begin{equation}
\label{eq:delta-Nc}
    \Delta_{N_c} = \dfrac{\vert\mathcal{R}_{\mathrm{med}}- \mathcal{R}_{\mathrm{med}}^{\mathrm{large}\text{-}N_c}\vert}{\vert\mathcal{R}_{\mathrm{med}}\vert}\, ,
\end{equation}
which vanishes when subleading color contributions are absent. We obtain the large-$N_c$ limit by applying two approximations: (i) setting $C_F=N_c/2$ and (ii) neglecting terms suppressed by inverse powers of $N_c$ in the evolution matrix $\tilde{\mathbb{M}}_{\sigma}^{\sigma'}$ from Eq.~\eqref{eq:schro-B}. The latter step leads to substantial simplifications in the numerical evaluation of $\mathcal{R}_{\mathrm{med}}$. To disentangle the effects of (i) and (ii), it is instructive to study the $z\to 0, 1$ endpoints. In these kinematic regions, oscillations between different color states vanish, as we analytically demonstrate in the Supplemental Material, meaning that $\Delta_{N_c}$ is entirely driven by $N_c$-corrections to $C_F$. For $g\to gg$ and $g\to q\bar q$ splittings the $z\to 0, 1$ limits of $\mathcal{R}_{\mathrm{med}}$ are proportional to $C_A$ and $C_F$, respectively. Consequently, $\Delta_{N_c}$ vanishes exactly for the former channel, but remains non-zero for the latter. For $q\to gq$ splittings, the behavior is asymmetric: the emission rate is dominated by $C_A$ in the $z\to 0$ limit (where $\Delta_{N_c}=0$), whereas it is governed by $C_F$ as $z\to 1$. At finite-$z$, $\Delta_{N_c}$ is non-zero across all splitting channels and its magnitude depends on the specific phase-space point, though it remains small throughout.

\begin{figure}[h] \centerline{%\hspace*{-.4cm}
\includegraphics[width=0.9\columnwidth]{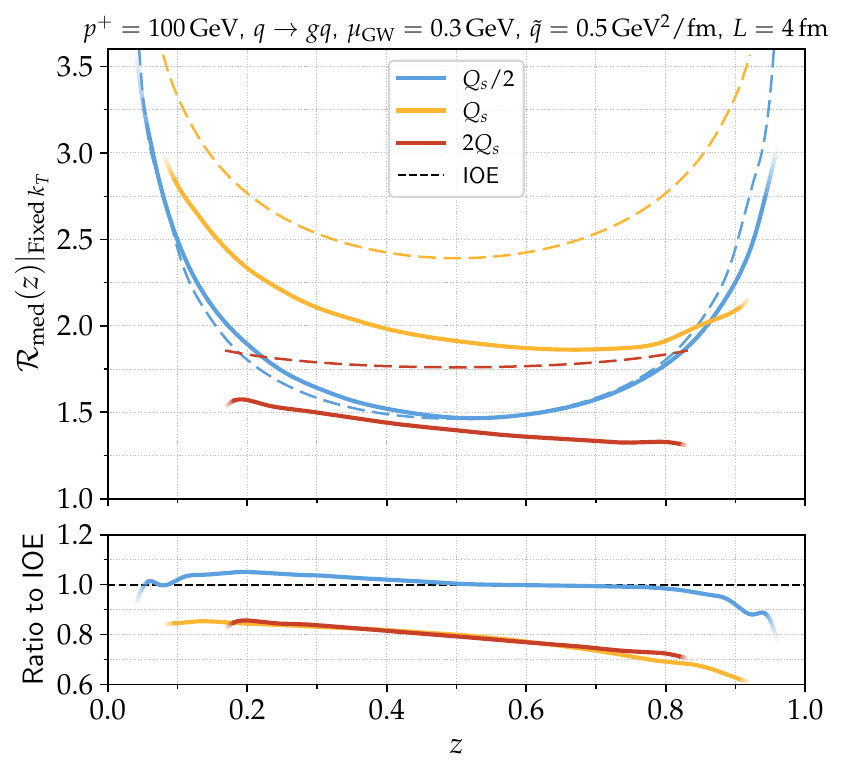}}
 \caption{Comparison between the large-$N_c$ limit of the numerical routine presented in this Letter and a semi-analytic approximation of ${\cal{R}}_{\rm med}(z,k_T)$ based on the Improved Opacity Expansion~\cite{Barata:2021wuf}.}
 \label{fig:ioe}
\end{figure}

Finally, we examine the magnitude of the discrepancies between the exact result for ${\cal{R}}_{\rm med}$ and standard approximations found in the literature, even though the multi-parametric nature of the problem precludes a definitive statement. Fig.~\ref{fig:ioe} compares our numerical results with a semi-analytic calculation based on the Improved Opacity Expansion (IOE) scheme at next-to-leading order~\cite{Barata:2021wuf}. This framework approximates the scattering potential according to Eq.~\eqref{eq:vba-coord-space} and relies on the $z\to 0$ approximation. Thus, the observed deviations of the IOE from the full numerical result reflect the sensitivity of ${\cal{R}}_{\rm med}$ to $\mathcal{O}(Q_s/\mu_\star)$ and finite-$z$ corrections. The impact of such terms is significant; for instance, they reach up to 20\% for $k_T=Q_s$ and $z\sim 0.5$, suggesting that the corrections to ${\cal{R}}_{\rm med}$ computed in this Letter could have a non-negligible phenomenological effect. 

In conclusion, we have developed a numerical framework to evaluate the single-emission differential cross-section in a dense QCD medium, which constitutes the fundamental building block for jet physics calculations in heavy-ion collisions. This framework establishes a new baseline for all QCD in-medium splitting functions by overcoming long-standing analytical simplifications, such as the large-$N_c$, soft, and harmonic-oscillator approximations. There is ample scope for phenomenological exploration using the medium-induced spectrum computed in this work. A natural next step is to study the impact of our exact results on jet observables with reduced sensitivity to non-perturbative corrections, such as groomed substructure observables~\cite{Mehtar-Tani:2016aco,Chien:2016led,Milhano:2017nzm,Caucal:2019uvr,Casalderrey-Solana:2019ubu,Caucal:2021cfb,Wang:2022yrp,JETSCAPE:2023hqn,Kudinoor:2025gao,Apolinario:2026hff}, and/or resummation effects, such as energy correlators~\cite{Andres:2022ovj,Barata:2023bhh,Barata:2023zqg,Yang:2023dwc,Fu:2024pic,Xing:2024yrb,Singh:2024vwb,Bossi:2024qho,Barata:2024ukm,Barata:2025uxp,Barata:2025zku,Kudinoor:2025gao,Andres:2025yls,Apolinario:2025vtx,Barata:2026pgh}. Additionally, we aim to integrate these splitting kernels into state-of-the-art jet quenching Monte Carlo event generators, thereby improving the modeling of parton showers in heavy-ion collisions~\cite{Zapp:2012ak,Casalderrey-Solana:2014bpa,JETSCAPE:2017eso,Caucal:2018dla}. Such studies are crucial for fully exploiting upcoming high-precision data from both sPHENIX~\cite{Belmont:2023fau} and the LHC experiments.

\begin{acknowledgments}
We are grateful to Carlota Andrés, João Barata and Fabio Dominguez for clarifying some important aspects in their work and for engaging in detailed comparisons, and to Rafael Soares for clarifications on the Faber polynomial expansion. We further thank Gregory Soyez for engaging in multiple discussions throughout this project. A.S.O.’s work was supported by the Ramón y Cajal program under grant RYC2022-037846-I and by ERDF (grant PID2024-161668NB-100). A.S.O. and J.G.M. were also supported by grant NSF PHY-2309135 to the Kavli Institute for Theoretical Physics (KITP).This work is part of a project
that has received funding from the European Research
Council(ERC) under the European Union’s Horizon 2020
research and innovation programme (Grant agreement
No. 835105, YoctoLHC).
We acknowledge support from Fundação para a Ciência e a Tecnologia
(FCT I.P.), under ERC-PT A-Projects ‘Unveiling’, financed
by PRR, NextGenerationEU (J.G.M. and M.L.), and contract 2024.00319.BD (M.L.).

\end{acknowledgments}

\bibliographystyle{apsrev4-2}
\bibliography{biblio}% Produces the bibliography via BibTeX.

\onecolumngrid
\newpage

% uncomment thisto start the supplemental material at page 1 (e.g. for
% PRL submission)
%\setcounter{page}{1}
\appendix
\makeatletter
\renewcommand\@biblabel[1]{[#1S]}
\makeatother

%======================================================================
\section*{Supplemental material}
\subsection{Medium description}
\label{app:medium}
A central ingredient in computing the splitting spectrum is the description of the QCD medium through which the splitting parton and its offsprings propagate. Following the standard assumptions of the BDMPS-Z formalism, we adopt a semi-classical treatment of the medium. Specifically, interactions between the hard partons and the medium are modeled via a background color field, $A$, acting over the light-cone time interval $[0,L]$. When computing the squared matrix element, we average over all possible field configurations assuming Gaussian statistics and locality in both color and light-cone time. Consequently, the only non-trivial averaging over $A$ reduces to the two-point correlator:  
\begin{equation}
\langle A^{-,a}(\bm{r},t) A^{-,b}(\bm{r}',t') \rangle = n(t) \delta^{ab}\delta(t-t')\gamma(\bm{r}-\bm{r}'),
\end{equation}
where $n(t)$ represents the density of scattering centers and $\gamma(\bm{x})$ is defined as
\begin{equation}
    \gamma(\bm{x}) = \int_{\bm{q}}e^{i\bm{q}\cdot\bm{x}} \, \dfrac{d \sigma}{d^2 \bm{q}}.
\end{equation}
Here, $d\sigma/d^2\bm{q}$ denotes the elastic $2\to 2$ differential cross-section for a single scattering between the hard probe and the medium. By introducing the dipole cross-section $\sigma(\bm{x}) = g^2(\gamma(0) - \gamma(\bm{x}))$, where $g^2\equiv 4\pi\alpha_s$, we can readily establish the connection to Eq.~\eqref{eq:in-out} via the process-dependent scattering potential $\tilde v_{ba}$. In transverse-coordinate representation, this potential reads  
\begin{equation}
    v_{ba}(\bm{x}) = n(t)\left[\dfrac{c_{cba}}{2}\sigma(\bm{x}) + \dfrac{c_{acb}}{2}\sigma(z\bm{x}) + \dfrac{c_{bac}}{2}\sigma((1-z)\bm{x})\right].
    \label{eq:vba}
\end{equation}
In the preceding expression, we have introduced the color factors  $c_{ijk} = C_i + C_j - C_k$, where $C_i$ denotes the Casimir of the color representation for particle $i$, which splits into the $jk$ pair. The results shown in the main text assumed the density of scattering centers to be constant, $n(t)=n$. However, our numerical routines can be adapted to time-dependent densities by considering a Magnus expansion of the time-ordered evolution operator in Eq. \eqref{eq:f-evol}, which is systematically improvable and does not affect stability. We leave the question of expanding media for future work. 

To complete the description of the medium, all that remains to be specified is the elastic scattering cross-section. In this work, we consider the two most widely used models in the literature, although our numerical routines are flexible enough to accommodate any arbitrary potential, such as those discussed in Ref.~\cite{Schlichting:2021idr}. First, the Gyulassy-Wang (GW) model~\cite{Wang:1992qdg} describes the medium as a collection of static, screened scattering centers, yielding    
\begin{equation}
\label{eq:gw-potential}
    \dfrac{d \sigma_{\mathrm{GW}}}{d^2\bm{q}} = \dfrac{g^4}{|\bm{q}^2 + \mu^2|^2}
\end{equation}
where $\mu$ serves as an infrared regulator. Alternatively, assuming the medium is in thermal equilibrium, the cross-section can be computed using Hard Thermal Loop (HTL) perturbation theory. This approach yields~\cite{Aurenche:2002pd}
\begin{equation}
\label{eq:htl-potential}
    \dfrac{d \sigma_{\mathrm{HTL}}}{d^2\bm{q}} = \dfrac{g^2 \mu_D^2 T}{\bm{q}^2(\bm{q}^2 + \mu^2)},
\end{equation}
where $\mu_D^2 = (1 + N_f/6)g^2T^2$ is the squared Debye mass. Both models can be recast into a unified form at leading power, given by
\begin{equation}
\label{app:eq:potential}
    n\sigma(\bm{x}) = \dfrac{1}{4}\tilde{q}\, \bm{x}^2\ln\dfrac{1}{\mu_\star\bm{x}^2} + \mathcal{O}(\mu_\star^2\bm{x}^4),
\end{equation}
where $\mu_\star$ corresponds to an effective screening mass. The explicit expression for $\mu_\star$ in terms of the parameters of each scattering model was derived in Ref.~\cite{Barata:2020rdn}. The second parameter entering Eq.~\eqref{app:eq:potential} is $\tilde{q}$, which is defined as
\begin{equation}
\tilde{q} = n g^2 \int^{Q}\frac{d^2q}{(2\pi)^2}q^2 \frac{d\sigma}{d^2 \bm{q}}\, ,
\end{equation}
where $Q$ denotes an ultraviolet cutoff and $\tilde{q} \equiv \hat q_R/C_R$, with $\hat q_R$ being the standard jet quenching parameter for a soft particle in color representation $R$. Finally, a standard approach in the literature, known as harmonic oscillator approximation, consists of neglecting the logarithm in Eq.~\eqref{app:eq:potential}, which reduces most subsequent integrals to a solvable Gaussian form.

\subsection{Explicit formulae for color-mixing matrices}
\label{app:colour-potential}
In this section, we provide explicit expressions for the color matrix $\tilde{\mathbb{M}}^{\sigma'}_\sigma$ that enters the evolution equation, Eq.~\eqref{eq:schro-B}, of the main text. For completeness, we also present the $\gamma\to q\bar{q}$ case, which was omitted in the main text. These results are derived within the framework established in Refs.~\cite{Isaksen:2020npj} and \cite{Isaksen:2023nlr}. 

The starting point is the definition of the splitting kernel ${\cal K}(t_2,t_1)$ and quadrupole ${\cal Q}(t_\infty,t_2)$ entering Eq.~\eqref{eq:full-spectrum} in terms of path-integrals over traces of Wilson lines, i.e.  
\begin{subequations}
 \begin{align}
    \mathcal{K}(\bm{u}_2, \bm{u}_1|t_2,t_1) = \int_{\bm{u}_1}^{\bm{u}_2} \mathcal{D}\bm{u} \, e^{i\int_{t_1}^{t_2} d s \frac{\omega}{2}\dot{\bm{u}}^2} \mathbb{C}^{(3)}(\bm{u})\, ,
    \label{eq:app-kernel}%
\end{align}   
\begin{align}
    \mathcal{Q}(\bm{u}, \bar{\bm{u}}, \bm{u}_2, \bar{\bm{u}}_2|t_\infty, t_2) = \int_{\bm{u}_2}^{\bm{u}} \mathcal{D}\bm{u}'\int_{\bm{\bar{u}}_2}^{\bm{\bar{u}}}\mathcal{D}\bm{\bar{u}}' \, e^{i\int_{t_2}^{t_\infty} d s \frac{\omega}{2}(\dot{\bm{u}}'^2 - \dot{\bm{\bar{u}}}'^2)}\mathbb{C}^{(4)}(\bm{u}', \bar{\bm{u}}')\, .
    \label{eq:app-quadrupole} %
\end{align}
\end{subequations}
While the main text is formulated in momentum space, here we switch to position space to facilitate comparisons with existing literature. The relations expressing the $(\bm{u},\bar{\bm{u}})$-coordinates in terms of the parton positions are provided in App. B and A of Refs. \cite{Blaizot:2012fh, Isaksen:2023nlr}, respectively. 

The correlators $\mathbb{C}^{(n)}$ are obtained by averaging traces of $n$ Wilson lines over the medium background field configurations. For $n\geq3$, $\mathbb{C}^{(n)}$ depends on the specific splitting process under consideration.  For the three-point function ($n=3$), the correlator reads
\begin{equation}
\label{eq:C3}
     \mathbb{C}^{(3)}(t_2, t_1 | \bm{u}) = \exp\left[-\int_{t_1}^{t_2}  d s \, v_{ba}(\bm{u}(s)) \right],
\end{equation}
where $v_{ba}(\bm{u})$ is defined in Eq.~\eqref{eq:vba}. For a generic dipole potential $v_{ba}$, the path-integral in Eq.~\eqref{eq:app-kernel} cannot be evaluated analytically. Consequently, it is more convenient to directly solve the evolution equation presented in the main text, see Eq.~\eqref{eq:in-out}. A notable exception is the harmonic approximation of the scattering potential discussed in the previous section, since then Eq.~\eqref{eq:app-kernel} reduces to Gaussian integrals and can be solved analytically.  

For $n=4$, we define the process-dependent correlators $\mathbb{C}^{(4)}$ in terms of traces of fundamental Wilson lines as follows: 
\begin{subequations}
\begin{align}
\mathbb{C}_{\gamma\to q\bar q}^{(4)} &= \frac{1}{N_c} \langle\mathrm{Tr}[V_1V_2^\dagger V_{\bar{2}} V_{\bar{1}}^\dagger]\rangle ,  \\ 
\mathbb{C}_{g \to q\bar q}^{(4)} &=  \frac{1}{N_c^2 -1} \langle\mathrm{Tr}[V_{\bar{1}}V_1^\dagger]\,\mathrm{Tr}[V_2V_{\bar{2}}^\dagger ]-\frac{1}{N_c}\mathrm{Tr}[V_1V_2^\dagger V_{\bar{2}} V_{\bar{1}}^\dagger]\rangle,\\ 
\mathbb{C}_{q\to gq}^{(4)} &= \frac{1}{N_c^2-1} \langle\mathrm{Tr}[V_1V_2^\dagger V_{\bar{2}} V_{\bar{1}}^\dagger]\,\mathrm{Tr}[V_2V_{\bar{2}}^\dagger ] - \frac{1}{N_c} \mathrm{Tr}[V_1V_{\bar{1}}^\dagger ] \rangle,
\\
\mathbb{C}_{g\to gg}^{(4)} &= \frac{1}{N_c(N_c^2-1)} 
\langle\mathrm{Tr}[V_{\bar{1}}V_1^\dagger]\,\mathrm{Tr}[V_2V_{\bar{2}}^\dagger ]\,\mathrm{Tr}[V_1V_2^\dagger V_{\bar{2}} V_{\bar{1}}^\dagger]  -  \mathrm{Tr}[V_{\bar{1}}V_2^\dagger V_{\bar{2}}V_{\bar{1}}^\dagger V_1 V_{\bar{2}}^\dagger V_2 V_1^\dagger] \rangle,
\end{align}
\label{eq:app-C-eqs}%
\end{subequations}
where $(V_i, V_{\bar{i}}) \equiv (V(\bm{r}_i), V(\bar{\bm{r}}_i))$, for $i\in \{1,2\}$. Here, $i = 1$ denotes the parton carrying an energy fraction $z$, and the bar notation $\bar{\bm{r}}$ indicates positions within the complex conjugate amplitude. Evaluating the path integrals in Eq.~\eqref{eq:app-quadrupole} with these correlators is highly non-trivial. Alternatively, one can derive a closed system of differential equations governing the evolution of each correlator. In the following, we briefly outline this approach. 

Let us consider pairs of fundamental Wilson lines denoted by $W_{a_i, b_i} \equiv V_{a_i}V_{b_i}^{\dagger}$. In general, a $2n$-point function can be expressed in the form
\begin{equation}
    C_p = \delta_{i_1i_2\dots i_n}^{p[j_1j_2\dots j_n]} (W_{a_1,b_1})_{i_1j_1} (W_{a_2, b_2})_{i_2j_2}\dots(W_{a_n, b_n})_{i_n j_n},
\end{equation}
where $\delta_{\alpha_1\dots\alpha_n}^{\beta_1\dots\beta_n} = \delta_{\alpha_1}^{\beta_1}\dots\delta_{\alpha_n}^{\beta_n}$ and $p[j_1j_2\dots j_n] =  j_{p_1}\dots j_{p_n}$ represents a permutation of the index sequence $[j_1j_2\dots j_n]$. We label each correlator by its associated permutation, writing $C_{(p_1\dots p_n)} \equiv C_p$. As shown in \cite{Isaksen:2020npj}, correlators of this class satisfy
\begin{equation}
    \frac{d}{dt}C_{p} = -\frac{1}{2}n(t)\left[\bm{M}_{\mathrm{diag}, p}\, C_p + \sum_{\pi=p  \,\mathrm{+ \,one \,swap}} (\bm{M}_{\mathrm{non-diag}})_{\pi}^{p}\,C_{\pi}\right],
\end{equation}
where the diagonal matrix elements are given by
\begin{equation}
    \bm{M}_{\mathrm{diag}, (p_1\dots p_n)} = \sum_{i=1}^n  \left[N_c\sigma_{b_ia_{p_i}}- \frac{1}{N_c}\sigma_{a_ib_i}\right] + \dfrac{1}{N_c}\ \sum_{i=1}^n \sum_{j>i}^n(\sigma_{a_ia_j} + \sigma_{b_ib_j} - \sigma_{b_ia_j} - \sigma_{a_ib_j}) 
\end{equation}
with $\sigma_{ab} = \sigma(\bm{r}_a - \bm{r}_b)$ defined in Eq.~\eqref{app:eq:potential}. The non-diagonal elements, which couple the permutation $p$ to configurations $\pi$ differing by a single index swap $k \leftrightarrow l$, e.g. $(p_1,...,p_k,...p_l,...p_n) \to \pi = (p_1,...,p_l,...p_k,...p_n)$, read
\begin{equation}
   (\bm{M}_{\mathrm{non-diag}})_{\pi}^{(p_1\dots p_n)} = \sigma_{b_ka_{p_l}} +  \sigma_{a_{p_k}b_l} -  \sigma_{a_{p_k}a_{p_l}} -  \sigma_{b_kb_l}\, .
\end{equation}

Using these ingredients, the evolution equations for the correlators in Eq. \eqref{eq:app-C-eqs} can be written in a compact matrix form as 
\begin{align}
    \dfrac{d\mathcal{C}_{\sigma}^{ijk}}{dt} = \mathbb{M}_{\sigma}^{\sigma'}\mathcal{C}_{\sigma'}^{ijk},
\end{align}
where the index $\sigma$ runs over all color configurations that couple to the correlator of interest. In what follows, we define $\mathcal{C}_1^{ijk} \equiv \mathbb{C}^{(4)}_{i\to jk}$, meaning that $\sigma=1$ corresponds directly to the physical correlators defined in Eq.~\eqref{eq:app-C-eqs}. These correlators are normalized such that the initial value is $\mathcal{C}_\sigma(0)=1$; for configurations where the initial value is zero, no normalization factor is applied. Furthermore, the evolution matrix $\mathbb{M}_{\sigma}^{\sigma'}$ depends only on a few linear combinations of the dipole scattering amplitudes: 
\begin{subequations}
\begin{align}
    \Sigma_{0} &= \sigma_{12} +\sigma_{\bar{1}\bar{2}} =    \sigma(\bm{u}) + \sigma(\bar{\bm{u}}),\\
    \Sigma_+ &= \sigma_{1\bar{1}} = \sigma(z(\bm{u} - \bar{\bm{u}})) \\
    \Sigma_- &= \sigma_{2\bar{2}} = \sigma((1-z)(\bm{u} - \bar{\bm{u}})) \\
    \Sigma_{zs} &= \sigma_{\bar{1}2} + \sigma_{1\bar{2}}  =  \sigma((1-z)\bm{u} + z\bar{\bm{u}}) + \sigma(z\bm{u} + (1-z)\bar{\bm{u}}).
\end{align}
\label{app:eq:sigma-def}%
\end{subequations}
We note that $\Sigma_0$ and $\Sigma_{zs}$ are invariant under the energy fraction exchange $z \to 1-z$, whereas $\Sigma_{\pm}$ transform into each other ($\Sigma_{+}\leftrightarrow\Sigma_{-}$). This transformation property ensures that the combination $\Sigma_{\pm} = \Sigma_{+}+ \Sigma_{-}$ remains invariant under the exchange of energy fractions. In what follows, we provide explicit expressions for each splitting channel. 

\begin{itemize} 
\item $\gamma \to q\bar{q}$: the evolution equations for this channel were previously derived in Ref. \cite{Isaksen:2020npj}. The correlator associated with this vertex stems from the double-pair structure 
\begin{equation}
\mathbb{C}_{\gamma\to q\bar q}^{(4)} = \frac{1}{N_c} C_{(21)} \quad{\rm with}\quad C_p = \delta_{i_1 i_2}^{p[j_1 j_2]} (V_1V_2^{\dagger})_{i_1 j_1}(V_{\bar{2}}V_{\bar 1}^{\dagger})_{i_2 j_2}\, .
\end{equation}
The resulting coupled system of equations reads
\begin{subequations}
\begin{align}
\frac{d\mathcal{C}^{\gamma q\bar q}_{1}}{dt}
&= \dfrac{n(t)}{2}\left[\Big(2C_F\Sigma_\pm + \frac{1}{N_c}(\Sigma_{zs} - \Sigma_0)\Big)\mathcal{C}^{\gamma q\bar q}_{1} - N_c(\Sigma_{zs} - \Sigma_0)\mathcal{C}^{\gamma q\bar q}_{2}\right] ,
\\
\frac{d\mathcal{C}^{\gamma q\bar q}_{2}}{dt}
&= \dfrac{n(t)}{2}\left[-\dfrac{1}{N_c}(\Sigma_{zs} - \Sigma_\pm)\mathcal{C}^{\gamma q\bar q}_{1} + \Big(2C_F\Sigma_0 + \frac{1}{N_c}(\Sigma_{zs} - \Sigma_\pm)\Big)\mathcal{C}^{\gamma q\bar q}_{2}\right],
\end{align}
\label{app:eq:gamma-correlator}%
\end{subequations}
where the auxiliary state is defined as $\mathcal{C}_ 2^{\gamma q\bar{q}} \equiv \frac{1}{N_c^2}C_{(12)}$. 

\item $g \to q\bar{q}$: the underlying correlator contains two independent color structures. We write 
\begin{equation}
\mathbb{C}_{g\to q\bar q}^{(4)} =  \frac{1}{N_c^2-1} \, [C_{(12)} - \frac{1}{N_c} \, D_{(21)}] \quad {\rm with} \quad
\begin{cases}
      C_{p} \equiv \delta_{i_1 i_2}^{p[j_1 j_2]} (V_{\bar{1}}V_1^{\dagger})_{i_1 j_1}(V_{2}V_{\bar 2}^{\dagger})_{i_2 j_2}  , \\
     \displaystyle
     D_p \equiv \delta_{i_1 i_2}^{p[j_1 j_2]}(V_1V_2^{\dagger})_{i_1 j_1}(V_{\bar{2}}V_{\bar 1}^{\dagger})_{i_2 j_2}.
  \end{cases} 
\end{equation} 
Although the full evolution initially yields a system of four coupled differential equations, they can be combined into a closed set of two equations that read
\begin{subequations}
\begin{align}
\frac{d\mathcal{C}^{g q\bar q}_{1}}{dt}
&= \dfrac{n(t)}{2}\left[\Big(2C_F\Sigma_\pm + \frac{1}{N_c}(\Sigma_{zs} - \Sigma_0)\Big) \mathcal{C}^{g q\bar q}_{1}+\dfrac{1}{N_c}\left(\Sigma
   _0-\Sigma _{zs}\right) \mathcal{C}^{g q\bar q}_{2}\right],
\\
\frac{d\mathcal{C}^{gq\bar q}_{2}}{dt}
&=  \dfrac{n(t)}{2}\left[\left(\Sigma _{\pm}-\Sigma _{zs}\right) \mathcal{C}^{g q\bar q}_{1} + \Big(2C_F\Sigma_0 + \frac{1}{N_c}(\Sigma_{zs} - \Sigma_\pm)\Big) \mathcal{C}^{gq\bar q}_{2}\right],
\end{align}
\label{app:eq:gqq-correlator}%
\end{subequations}
where $\mathcal{C}^{gq\bar q}_{2} \equiv N_c C_{(21)} - D_{(12)}$  satisfies $\mathcal{C}_2^{gq\bar{q}}(t=0) = 0$. As a consequence, its associated source term in the evolution equation for $\cal{B}$ vanishes, i.e., $\xi_{\sigma = 2} = 0$ in Eq. \eqref{eq:schro-B}. 

\item $q\to gq$: the physical correlator is expressed as 
\begin{equation}
\mathbb{C}^{(4)}_{q\to gq} = \frac{1}{N_c^2-1} [C_{(213)} - \frac{1}{N_c}C_{(312)}] \quad{\rm with}\quad  C_p = \delta_{i_1 i_2 i_3}^{p[j_1 j_2j_2]} (V_1V_2^\dagger)_{i_1j_1}(V_{\bar{2}}V_{\bar{1}}^\dagger)_{i_2j_2}(V_2V_{\bar 2}^\dagger)_{i_3j_3}\, .
\end{equation}
The resulting reduced system of evolution equations is given by
\begin{subequations}
\begin{align}
\frac{d\mathcal{C}_1^{qgq}}{dt}
= &
\dfrac{n(t)}{2} \left[
2\left(
C_F\Sigma_+
+ N_c \,  \Sigma_{-}
\right)\mathcal{C}_1^{qgq} - N_c\left(
\Sigma_{zs}  - \Sigma_0  
\right)\mathcal{C}^{qgq}_2\right] \, ,
\\
\frac{d\mathcal{C}_2^{qgq}}{dt} 
= &
\dfrac{n(t)}{2}\left[\left(
-\frac{\Sigma_+}{N_c}
+ N_c(\Sigma_0+\Sigma_-)
\right)\mathcal{C}^{qgq}_2
+\dfrac{1}{N_c^2(N_c^2 -1)}\left(
\Sigma_{zs} - \Sigma_0
\right)\mathcal{C}^{qgq}_3\right] \, ,
\\
\frac{d\mathcal{C}^{qgq}_3}{dt}
= &
\dfrac{n(t)}{2}\big[N_c^2(N_c^2-1)\big(\left(
\Sigma_{zs} + \Sigma_0 - 2\Sigma_{\pm}
\right)\mathcal{C}^{qgq}_1
+\left(
\Sigma_{zs} - \Sigma_0
\right)\mathcal{C}^{qgq}_2\big)
\nonumber \\ 
&+\left(
N_c(\Sigma_{zs}+\Sigma_-) -\frac{\Sigma_+}{N_c}
\right)\mathcal{C}^{qgq}_3\big],
\end{align}
\end{subequations}
where the new color configurations read
\begin{subequations}
\begin{align}
\mathcal{C}^{qgq}_2 &= \dfrac{1}{N_c(N_c^2-1)}[C_{(123)} - C_{(231)}] \, ,
  \\
\mathcal{C}^{qgq}_3 &= C_{(132)}-C_{(213)}-N_c C_{(231)} + C_{(321)}.
\end{align}
\label{app:eq:qgq-correlator}%
\end{subequations}
In this case, $\mathcal{C}^{qgq}_3(t=0) = 0$ and $\xi_{\sigma=3} = 0$. Notably, the potential matrix contains terms proportional to $\Sigma_{+}$ and $\Sigma_{-}$. Their origin can be traced back to the lack of symmetry of $q\to gq$ splittings under $z\to 1-z$ transformations.

\item $g\to gg$: the evolution equations for this vertex had not been explicitly reported in previous literature. The color structures associated with this splitting channel are 
\begin{equation}
\mathbb{C}^{(4)}_{g\to gg} = \frac{1}{N_c(N_c^2-1)} [C_{(1243)} - D_{(2341)}] \quad{\rm with}\quad
\begin{cases}
      C_{p} \equiv \delta^{p[j_1j_2j_3j_4]}_{i_1i_2i_3i_4}(V_{\bar{1}}V_1^{\dagger})_{i_1 j_1}(V_ 2V_{\bar 2}^{\dagger})_{i_2 j_2}(V_1V_2^{\dagger})_{i_3 j_3}(V_{\bar{2}}V_{\bar 1}^{\dagger})_{i_4 j_4} , \\
     \displaystyle
     D_p \equiv \delta^{p[j_1j_2j_3j_4]}_{i_1i_2i_3i_4} (V_{\bar{1}}V_2^{\dagger})_{i_1 j_1}(V_{\bar 2}V_{\bar 1}^{\dagger})_{i_2 j_2}(V_1V_{\bar 2}^{\dagger})_{i_3 j_3}(V_2V_{1}^{\dagger})_{i_4 j_4}.
  \end{cases} 
\end{equation}
Individually, each structure generates a system of $24$ coupled differential equations, leading to a total of $48$ equations. Remarkably, this full set can be reduced to only six independent equations, which read: 
\begin{subequations}
\begin{align}
    \frac{d\mathcal{C}_1^{ggg}}{dt} & = \dfrac{n(t)}{2}\Big[ \, 2  N_c \Sigma _{\pm}\mathcal{C}_1^{ggg}+ N_c\left(\Sigma _0-\Sigma
   _{zs}\right)\mathcal{C}_2^{ggg}\Big] \, ,\\
   %%%%
   \frac{d\mathcal{C}_2^{ggg}}{dt} & = \dfrac{n(t)}{2}\Big[\dfrac{2}{N_c}  \left(\Sigma _0-\Sigma _{zs}\right)\mathcal{C}_1^{ggg}+ N_c
   \left(\Sigma _{\pm}+\Sigma _0\right)\mathcal{C}_2^{ggg} -\dfrac{2}{N_c} \left(\Sigma _0-\Sigma _{zs}\right)\mathcal{C}_3^{ggg}\Big]\, , \\
   %%%%
   \frac{d\mathcal{C}_3^{ggg}}{dt} &= \dfrac{n(t)}{2}\Big[- N_c \left(\Sigma _{\pm}-\Sigma _0\right)\mathcal{C}_1^{ggg} - N_c
   \left(\Sigma _{\pm}-\Sigma _{zs}\right)\mathcal{C}_2^{ggg}+N_c \left(\Sigma _{\pm}+\Sigma
   _{zs}\right) \mathcal{C}_3^{ggg} \\
   &-\dfrac{2}{N_c(N_c^2-1)}\left(-2 \Sigma _{\pm}+\Sigma _0+\Sigma _{zs}\right) \mathcal{C}_4^{ggg} - 3
   N_c \left(\Sigma _{\pm}-\Sigma _{zs}\right)\mathcal{C}_5^{ggg}\Big]\, , \nonumber\\
   %%%%%
    \frac{d\mathcal{C}_4^{ggg}}{dt} &= \dfrac{n(t)}{2}\Big[ 2N_c(N_c^2-1)\left(\Sigma_0 - \Sigma_{\pm}\right)\mathcal{C}_1^{ggg} + 2 N_c^3(N_c^2-1) \left(\Sigma_{\pm} - \Sigma_{zs}\right)\mathcal{C}_2^{ggg} \nonumber\\
   &-2N_c(N_c^2-1)\left(\Sigma_0 - \Sigma_{\pm}\right)\mathcal{C}_3^{ggg}  \nonumber  - N_c \left(\Sigma_{\pm} - 3\Sigma_{zs}\right)\mathcal{C}_4^{ggg} \\
   &+ 2 N_c^3(N_c^2-1) \left(\Sigma_{\pm} + 2\Sigma_0 - 3\Sigma_{zs}\right)\mathcal{C}_5^{ggg}
   - N_c^3(N_c^2-1) \left(\Sigma_{\pm} + 4\Sigma_0 - 5\Sigma_{zs}\right)\mathcal{C}_6^{ggg} \Big] \, ,\\
   %%%%%%%%
   \frac{d\mathcal{C}_5^{ggg}}{dt}  &= \dfrac{n(t)}{2}\Big[N_c\left(\Sigma_{\pm} - \Sigma_{zs}\right)\mathcal{C}_2^{ggg} + \dfrac{1}{N_c(N_c^2-1)}\left(\Sigma_{zs} - \Sigma_0\right)\mathcal{C}_4^{ggg} \\
   &- 2 N_c \left(\Sigma_{zs} - 2\Sigma_0\right)\mathcal{C}_5^{ggg}  - 3 N_c \left(\Sigma_0 - \Sigma_{zs}\right)\mathcal{C}_6^{ggg}\Big] \nonumber \, ,\\ 
   %%%%%%%
   \frac{d\mathcal{C}_6^{ggg}}{dt} &= \dfrac{n(t)}{2}\Big[\dfrac{1}{N_c(1-N_c^2)}\left(\Sigma_0 - \Sigma_{\pm}\right)\mathcal{C}_4^{ggg} - 2 N_c \left(\Sigma_{\pm} - \Sigma_0\right)\mathcal{C}_5^{ggg} + N_c \left(\Sigma_{\pm} - \Sigma_0 + 2\Sigma_{zs}\right)\mathcal{C}_6^{ggg}\Big] \, .
\end{align}
\label{app:eq:ggg-correlator}%
\end{subequations}
The auxiliary color configurations are defined as
\begin{subequations}
\begin{align}
 %%%%%%%%%%%%%%
    \mathcal{C}_2^{ggg} &= \dfrac{1}{N_c^2(N_c^2-1)}\left[C_{(1234)} - C_{(1342)} + C_{(2143)} - C_{(4213)} + D_{(1342)} - D_{(2314)} + D_{(2431)} - D_{(3241)}\right] \, ,\\
    %%%%%%%%%%%%%%
  \mathcal{C}_3^{ggg} &= \dfrac{1}{2N_c(N_c
  ^2-1)}\Big[-C_{(1432)} + 2 C_{(2134)} - C_{(2341)} - C_{(2413)} - C_{(4231)} + 2 C_{(4312)} + D_{(1324)} - 2 D_{(1432)} + D_{(2134)} \nonumber \\
 &+N_c \left(C_{(1342)} + C_{(4213)} - D_{(1342)} - D_{(2431)} +  D_{(3142)} - 2 D_{(3214)} + D_{(3421)}\right)\Big] \, ,\\
  %%%%%%%%%%%%%%
  \mathcal{C}_4^{ggg} &=  -C_{(1342)} + C_{(2314)} + C_{(4132)} - C_{(4213)} - D_{(1234)} + D_{(1342)} + D_{(2431)} - D_{(3412)} \nonumber\\
  & + N_c \left[2 C_{(2134)} - C_{(4312)} + D_{(1432)} - 2 D_{(3214)}\right]\, , \\
%%%%%%%%%%%%%%
  \mathcal{C}_5^{ggg} &= \dfrac{1}{N_c(N_c^2-1)}\Big[C_{(2134)} - C_{(4312)} + D_{(1432)} - D_{(3214)} \Big] \, ,\\
%%%%%%%%%%%%%%
  \mathcal{C}_6^{ggg} &= \dfrac{1}{N_c(N_c^2-1)}\Big[D_{(1432)} - C_{(4312)}  \Big] \, .
\end{align}
\end{subequations}
Finally, at the initial condition, we find $\mathcal{C}_4^{ggg}(t=0) = 0$, which consequently yields $\xi_{\sigma = 4} = 0$.
\end{itemize}

The systems of equations presented above substantially simplify in two physical limits of interest: the soft emission limit ($z\to 0$) and the large-$N_c$ limit ($N_c\gg 1$). In what follows, we report explicit expressions for the evolution equations per splitting channel in each regime individually. 

\paragraph{Large-$N_c$ limit.} In this approximation, the dimensionalities of the systems reduce significantly: we find a single equation for the $g\to q\bar q$ channel, and only two independent equations for both $q\to gq$ and $g\to gg$ splittings. Conversely, there is no system size reduction for the $\gamma\to q \bar q$ vertex. Taking $N_c\gg 1$ in each splitting channel yields:
\begin{itemize}
\item $\gamma \to q\bar q$: 
\begin{subequations}
\begin{align}
\frac{d\mathcal{C}_1^{\gamma q\bar q}}{dt}
& \stackrel{N_c \gg 1}{\simeq}
\dfrac{N_c \, n(t)}{2}\left[
\Sigma_{\pm}\mathcal{C}_1^{\gamma q\bar q} + \left(
 \Sigma_0 -\Sigma_{zs}
\right)\mathcal{C}^{\gamma q\bar q}_2\right] \, ,
\\
\frac{d\mathcal{C}_2^{\gamma q\bar q}}{dt}
&\stackrel{N_c \gg 1}{\simeq}
\dfrac{N_c \, n(t)}{2}
\, \Sigma_0 \,
\mathcal{C}^{\gamma q\bar q}_2\, .
\end{align}
\label{app:eq:gamma-large-Nc}
\end{subequations}
\item $g\to q\bar{q}$:
\begin{equation}
\frac{d\mathcal{C}_1^{g q\bar q}}{dt}
 \stackrel{N_c \gg 1}{\simeq}
\dfrac{N_c}{2} \, n(t) \,
\Sigma_{\pm}\mathcal{C}_1^{g q\bar q}\, .
\label{app:eq:gqq-large-Nc}
\end{equation}
\item $q\to gq$:
\begin{subequations}
\begin{align}
\frac{d\mathcal{C}_1^{qgq}}{dt}
& \stackrel{N_c \gg 1}{\simeq}
\dfrac{N_c \, n(t)}{2}\left[
\left(
\Sigma_+
+ 2 \,  \Sigma_{-}
\right)\mathcal{C}_1^{qgq} + \left(
 \Sigma_0 -\Sigma_{zs}
\right)\mathcal{C}^{qgq}_2\right]\, ,
\\
\frac{d\mathcal{C}_2^{qgq}}{dt}
&\stackrel{N_c \gg 1}{\simeq}
\dfrac{N_c \, n(t)}{2}
(\Sigma_0+\Sigma_-)
\mathcal{C}^{qgq}_2\, .
\end{align}
\label{app:eq:qgq-large-Nc}
\end{subequations}
\item $g\to gg$:
\begin{subequations}
\begin{align}
    \frac{d\mathcal{C}_1^{ggg}}{dt} & \stackrel{N_c \gg 1}{\simeq}  \, \dfrac{N_c \,n(t)}{2}\Big[ \, 2   \Sigma _{\pm}\mathcal{C}_1^{ggg}+\left(\Sigma _0-\Sigma
   _{zs}\right)\mathcal{C}_2^{ggg}\Big] \, ,\\
   %%%%
   \frac{d\mathcal{C}_2^{ggg}}{dt} & \stackrel{N_c \gg 1}{\simeq} \, \dfrac{N_c\,n(t)}{2}
   \left(\Sigma _{\pm}+\Sigma _0\right)\mathcal{C}_2^{ggg}.
\end{align}
\label{app:eq:ggg-large-Nc}
\end{subequations}
\end{itemize}
The size of $N_c$-corrections for each splitting channel is quantified in Fig.~\ref{fig:nc}.

\paragraph{Soft limit.} When $z\to 0$, Eqs.~\eqref{app:eq:sigma-def} reduce to $\Sigma_{zs} \to \Sigma_0$, $\Sigma_+ \to 0$ and $\Sigma_{-}\to \Sigma_d$. Therefore, across all vertices, $\mathcal{C}_1$ decouples from the remaining color states. This yields a single evolution equation for each splitting channel:
\begin{subequations}
\begin{align}
\frac{d\mathcal{C}_1^{\gamma q\bar{q}}}{dt}
& \stackrel{z\to0 }{\simeq}
C_F \, n(t) \,
\Sigma_d \,
\mathcal{C}_1^{\gamma q\bar{q}} \, ,
\\
\frac{d\mathcal{C}_1^{gq\bar{q}}}{dt}
& \stackrel{z\to0 }{\simeq}
C_F  \, n(t) \,
\Sigma_d \, 
\mathcal{C}_1^{gq\bar{q}}\, ,
\\
\frac{d\mathcal{C}_1^{qgq}}{dt}
& \stackrel{z\to0 }{\simeq}
N_c \, n(t) \,
\Sigma_d \,
\mathcal{C}_1^{qgq} \, ,\\
\frac{d\mathcal{C}_1^{ggg}}{dt}  & \stackrel{z\to0 }{\simeq}  \, N_c \, n(t)\,\Sigma _{d}\, \mathcal{C}_1^{ggg} \, .
\end{align}
\label{app:eq:C1-soft-limit}%
\end{subequations}
We observe that, in this regime, the pairs of vertices $\{q \to gq,\, g \to gg\}$ and $\{\gamma \to q\bar{q},\, g \to q\bar{q}\}$ share the same equations, meaning that they share the quadrupole structure. Furthermore, their splitting kernels, governed by the medium potential $v_{ba}$ in Eq.~\eqref{eq:vba}, become identical within each pair. As a result, the medium-induced radiation spectra coincide within each pair, as the modifications are mainly imprinted in the soft particle. The evolution equations for $\mathcal{C}_1$ are identical when $z \to 1$ except in the $q \to gq$ channel, where the color factor becomes $C_F$, indicating that soft-quark modifications become dominant. Finally, the joint large-$N_c$ and soft limit is straightforwardly recovered by implementing the standard replacement $C_F \to N_c/2$ in Eq.~\eqref{app:eq:C1-soft-limit}.

%----------------------------------------------------------------------
\subsection{Considerations regarding the evolution equations of $\mathcal{A}$ and $\mathcal{B}$}
\label{app:evol-equations}
In this section we derive the evolution equations for the objects $\mathcal{A}$ and $\mathcal{B}$, demonstrating how to reduce their dimensionality by exploiting symmetries.

We begin with the position-space evolution for the splitting kernel $\mathcal{K}$~\cite{Isaksen:2023nlr},
\begin{equation}
    i\partial_t \mathcal{K}(\bm{u}, \bm{u}_1| t, t_0)  + \left(\dfrac{\nabla^2_{\bm{u}}}{2\omega} + iv_{ab}(\bm u) \right)\mathcal{K}(\bm{u}, \bm{u}_1|  t, t_0) = i\delta^{(2)}_{\bm{u} - \bm{u}_1} \delta(t - t_0).
    \label{app:eq:schro_kernel}
\end{equation}
The object $\mathcal{A}$, defined in Eq.~\eqref{eq:in-out}, is extracted via the relation
\begin{equation}
    \mathcal{A}(t,\bm{p}') = \int_0^td t_0 \, \int_{\bm{u}} e^{-i\bm{u}\cdot \bm{p'}} \, \nabla_{\bm{u}_1}\,  \mathcal{K}(\bm{u}, \bm{u}_1|  t, t_0) |_{\bm{u}_1 = 0}.
    \label{app:eq:a_deff}
\end{equation}
Combining this definition with Eq.~\eqref{app:eq:schro_kernel}, yields a momentum-space, Schrödinger-like equation for $\cal A$:
\begin{equation}
    i\partial_t  {\cal A}(t, \, \bm{{p^\prime}}) = \dfrac{{\bm{p^\prime}}^2}{2 \omega}{\cal A}(t, \bm{p^\prime}) - i\int_{\bm{q}}  \tilde{v}_{ba}(\bm{q}) \mathcal{A}(t, \bm{p}'-\bm{q}) -\bm{p}',
    \label{app:eq:schro-A}
\end{equation}
where $\tilde{v}_{ba}(\bm{q}) = \int_{\bm{u}} v_{ba}(\bm{u}) e^{-i\bm{u}\cdot \bm{q}}$, and the initial condition is $\mathcal{A}(t = 0) = \bm{0}$. This expression matches Eq.~\eqref{eq:schro-A}.

By adopting polar coordinates $\bm{p}= (p, \theta)$ and decomposing $\mathcal{A}$ into radial and angular components, $\mathcal{A}(t,\bm{p}) = a_p(t,p,\theta)\hat{\bm{p}} + a_\theta(t,p,\theta)\hat{\bm{\theta}}$, the rotational invariance of the medium potential, i.e. $\tilde{v}_{ba}(\bm{q})=\tilde{v}_{ba}(|\bm{q}|)$, ensures that the two components decouple. Given that ${\cal A}$ vanishes at $t=0$, the angular amplitude $a_\theta$ remains zero throughout the evolution. Furthermore, the scalar amplitude becomes purely isotropic $a_p(t,p,\theta)\equiv a(t,p)$, which reduces Eq.~\eqref{app:eq:schro-A} to a scalar $(1+1)$-dimensional equation,
\begin{equation}
    i\partial_t  a(t, \, p) = \dfrac{p^2}{2 \omega}a(t, \, p) -i \, n(t)\sum_a c_a \int_0^{\infty} d\tilde{p}\int_0^{2\pi} d\tilde{\theta} \, \tilde{p}\, V(\tilde{p})\left[a(t, p) - \frac{a(t,R_a)}{R_a}(p - \tilde{p} g_a(z)\cos\tilde{\theta})\right]- p, 
    \label{eq:schro_source_pol}
\end{equation}
where $c_a$ and $g_a(z)$ refer to the color and $z$-dependent factors (taking values $1$, $z$ or $1-z$) in Eq.~\eqref{eq:vba}, and $R_a \equiv \sqrt{p^2 + g_a(z)^2\tilde{p}^2 - 2p\tilde{p}g_a(z)\cos \tilde \theta}$. Finally, the isotropic potential is defined via the differential cross-section as $V(\bm{p}) = \frac{d\sigma}{d^2\bm{p}} \equiv V(p)$, with some examples provided by GW~\eqref{eq:gw-potential} and HTL~\eqref{eq:htl-potential} models.

Turning to the object $\mathcal{B}$, the starting point is the evolution equation for the quadrupole $\mathcal{Q}_{\sigma} =  \mathcal{Q}_\sigma(\bm{u}, \bar{\bm{u}}, \bm{u}_2, \bar{\bm{u}}_2|t, t_2)$, defined in Eq.~\eqref{eq:app-quadrupole}~\cite{Isaksen:2023nlr}:
\begin{equation}
    i\partial_t \mathcal{Q}_{\sigma}  + \left[\dfrac{\nabla_{\bm{u}}^2-\nabla^2_{\bm{\bar{u}}}}{2\omega}\delta_{\sigma}^{\sigma'} - i\mathbb{M}_{\sigma}^{\sigma'}(\bm{u}, \bm{\bar{u}}) \right]\mathcal{Q}_{\sigma'} = i\delta^{(2)}_{\bm{u} - \bm{u}_2}\delta^{(2)}_{\bm{ \bar{u}} - \bm{\bar{{u}}}_2}\delta(t - t_ 2) \, \xi_{\sigma}\, .
    \label{eq:quad}
\end{equation}
The functions $\mathcal{B}_{\sigma}$, defined in Eq.~\eqref{eq:in-in}, are related to the quadrupole $\mathcal{Q}_{\sigma}$ via the transformation
\begin{align}
\mathcal{B}_{\sigma}(t, \bm{k}, \bm{l}) =\int_{\bm{u}, \bar{\bm{u}}}e^{-i[\bm{u}\cdot(\bm{k}+\bm{l}) + \bar{\bm{u}}\cdot(\bm{l}-\bm{k})]}\int_0^t d t_1 \int_{t_1}^t d t_2 \int_{\bm{u}_2} (\nabla_{\bm{u}_1} \cdot \nabla_{\bar{\bm{u}}_2})  \, \mathcal{Q}_{\sigma}(\bm{u}, \bm{\bar{u}},\bm{u}_ 2, \bm{\bar{u}}_2|t, t_2)\,\mathcal{K}(\bm{u}_2, \bm{u}_1|t_2, t_1)|_{\bm{u}_1 = \bm{\bar{u}}_2 = \bm{0}}\, .
\end{align}
Differentiating this definition with respect to time and inserting Eq.~\eqref{eq:quad}, leads to the momentum-space evolution for $\mathcal{B}_{\sigma}$:
\begin{align}
    i\partial_t  {\cal B}_\sigma(t, \, \bm{k^\prime},\,\bm{l^\prime}) = &   \left[
        \,\frac{2\bm{k^\prime}\!\cdot\!\bm{l^\prime}}{\omega}\,
        \delta_{\sigma}^{\sigma'}
        + i\left(\tilde{\mathbb{M}}_{\sigma}^{\sigma'}  \ast \cdot \right)
    \right]
    \mathcal{B}_{\sigma'}\!\left(t, \bm{k^\prime},\,\bm{l^\prime}\right)\nonumber  \\
    &+(\bm{k^\prime} - \bm{l^\prime}) \cdot 
   \int_0^td t_1 \, \int_{\bm{u}_2} e^{-i\bm{u}_2\cdot (\bm{k'+l'})} \, \nabla_{\bm{u}_1}\,  \mathcal{K}(\bm{u}_2, \bm{u}_1| t, t_1) |_{\bm{u}_1 = 0}
    \, \xi_{\sigma} \, ,
    \label{eq:schro-B-app}
\end{align}
Noting that the integral in the second line, which we refer to as source term, corresponds to $\mathcal{A}(\bm{k}+\bm{l})$ allows us to recover Eq.~\eqref{eq:schro-B}. To reduce the dimensionality of Eq.~\eqref{eq:schro-B-app} to (3+1)-dimensions we proceed as follows. First, by exploiting the previously established isotropy of $\mathcal{A}$, i.e. $\mathcal{A}(t,\bm{p}) = a(t,p)\hat{\bm{p}}$, the scalar component of the source term simplifies to
\begin{equation}
\label{app:eq:source-term}
    S(\bm{k},\bm{l})\equiv(\bm{k} - \bm{l}) \cdot \mathcal{A}(\bm{k}+\bm{l}) = \dfrac{\bm{k}^2 - \bm{l}^2}{|\bm{k}+\bm{l}|}\,a(|\bm{k}+\bm{l}|)\, .
\end{equation}
Upon transforming to polar coordinates, $\bm{k} = (k, \phi)$ and $\bm{l} = (l, \eta)$, Eq.~\eqref{app:eq:source-term} becomes
\begin{equation}
    S(k, l, \psi\equiv \phi-\eta) = \dfrac{k^2 - l^2}{\sqrt{k^2+l^2+2kl\cos\psi}}\,a(\sqrt{k^2+l^2+2kl\cos\psi}),
\end{equation}
where $\psi$ denotes the relative angle between the two momentum vectors. This simplification reduces the dependence of  $S(\bm{k},\bm{l})$ to only three kinematic variables: $k, l$ and  $\psi$. The two remaining terms in Eq.~\eqref{eq:schro-B-app} also obey the same functional dependence. For instance, the kinetic term is a scalar product of two vectors, which we can naturally write in terms of the same three kinematic variables $k$, $l$ and $\psi$. The convolution term takes the general form 
\begin{equation}
    \tilde{\mathbb{M}}_{\sigma}^{\sigma'}\ast \mathcal{B}_{\sigma'} =  \sum_{\sigma', i} M^i_{\sigma \sigma'}\int_{\tilde{\bm{p}}}V(|\tilde{\bm{p}}|) \ [\mathcal{B}_{\sigma'}(\bm{k}, \bm{l})-\mathcal{B}_{\sigma'}(\bm{k} -a^i(z)\, \bm{\tilde{p}}, \,  \bm{l} -b^i(z)\,\bm{\tilde{p}})],
    \label{app:eq:convol}
\end{equation}
where we have omitted the temporal dependence and the index $i$ runs over $\{\tilde\Sigma_0, \tilde\Sigma_+, \tilde\Sigma_-,\tilde\Sigma_{zs}\}$, which correspond to the momentum-space representation of Eq.~\eqref{app:eq:sigma-def}. More concretely, the specific action of each of the components entering the sum in Eq.~\eqref{app:eq:convol} on a generic test function $f({\bm k'},{\bm l'})$ reads
\begin{subequations}
\begin{align}
    (\tilde\Sigma_0 \ast f)(\bm{k}, \bm{l}) & = \int_{\tilde{\bm p}}V(\tilde{\bm{p}}) \left[2\, f(\bm{k}, \bm{l}) -  f(\bm{k} - \tilde{\bm{p}}/2, \bm{l}-\tilde{\bm{p}}/2) - f(\bm{k} + \tilde{\bm{p}}/2, \bm{l}-\tilde{\bm{p}}/2) \right], \\
    (\tilde\Sigma_+ \ast f)(\bm{k}, \bm{l}) & = \int_{\tilde{\bm p}}V(\tilde{\bm{p}}) \left[\, f(\bm{k}, \bm{l}) -  f(\bm{k} - z\tilde{\bm{p}}, \bm{l}) \right], \\
    (\tilde\Sigma_- \ast f)(\bm{k}, \bm{l}) & = \int_{\tilde{\bm p}}V(\tilde{\bm{p}}) \left[\, f(\bm{k}, \bm{l}) -  f(\bm{k} - (1-z)\tilde{\bm{p}}, \bm{l}) \right], \\
    (\tilde\Sigma_{zs} \ast f)(\bm{k}, \bm{l}) & = \int_{\tilde{\bm p}}V(\tilde{\bm{p}}) \left[2\, f(\bm{k}, \bm{l}) -  f(\bm{k} - (z-1/2)\tilde{\bm{p}}, \bm{l}-\tilde{\bm{p}}/2) - f(\bm{k} + (z-1/2)\tilde{\bm{p}}, \bm{l}-\tilde{\bm{p}}/2) \right].
\end{align}
\end{subequations}
Let $R_{ik}\equiv|\bm{k} -a^i(z) \bm{\tilde{p}}|$ and $R_{il}\equiv|\bm{l} -b^i(z) \bm{\tilde{p}}|$ and let $\Psi$ represent the relative angle between these two shifted vectors. Exploiting the rotation symmetry of the medium, we can choose a coordinate system where the azimuthal orientation of $\bm{k}$ is fixed to $\phi = 0$. Under this geometric choice, the multi-dimensional convolution term in Eq.~\eqref{app:eq:convol} can be written in polar coordinates as 
\begin{equation}
    (\tilde{\mathbb{M}}_{\sigma}^{\sigma'}\ast \mathcal{B}_{\sigma'})(k,l,\psi) =  \sum_{\sigma', i} M^i_{\sigma \sigma'}\int_0^{2\pi} d\tilde{\theta} \int_0^{\infty} d\tilde{p} \, \tilde{p} \, V(\tilde{p}) \ [\mathcal{B}_{\sigma'}(k, l, \psi) - \mathcal{B}_{\sigma'}(R_{ik}, R_{il}, \Psi)]\, ,
\end{equation}
where $\bm{\tilde p} = (\tilde p, \tilde \theta)$. As anticipated, we observe that the final form of the convolution term only depends on three variables. 

\subsection{Results for $g\to gg$ and $g\to q\bar q$}
\begin{figure}[t]
\includegraphics[width=0.49\textwidth]{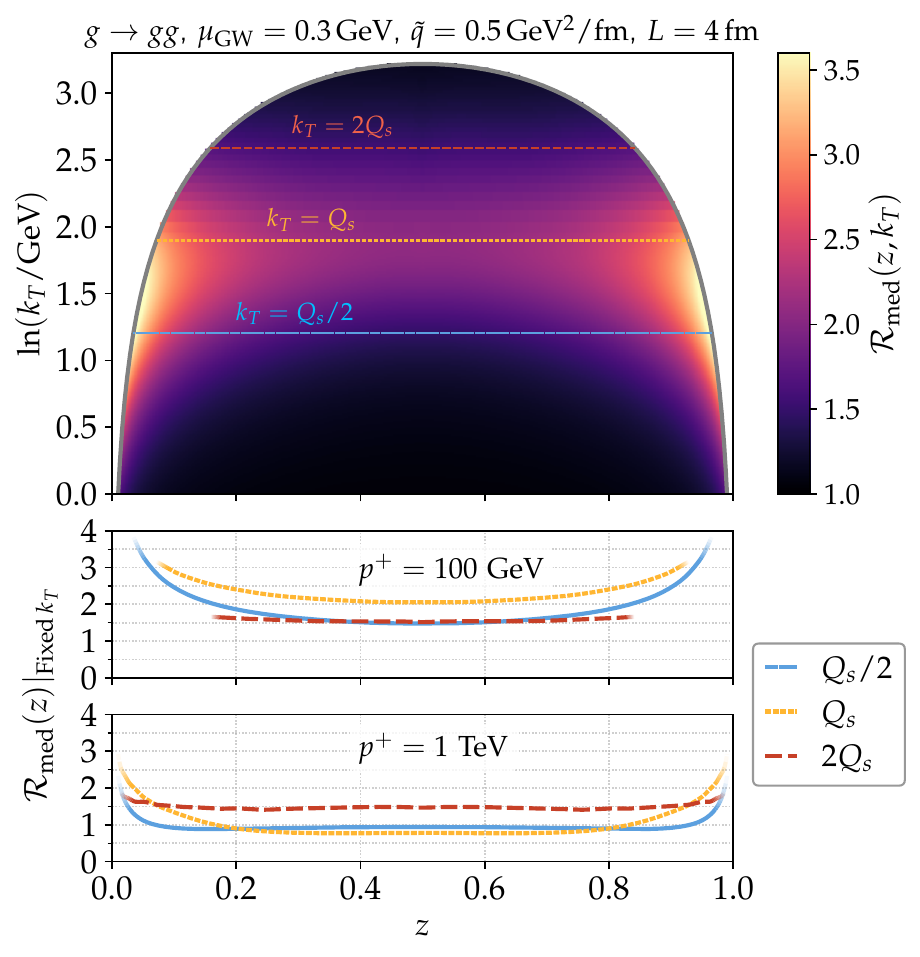}
\includegraphics[width=0.49\textwidth]{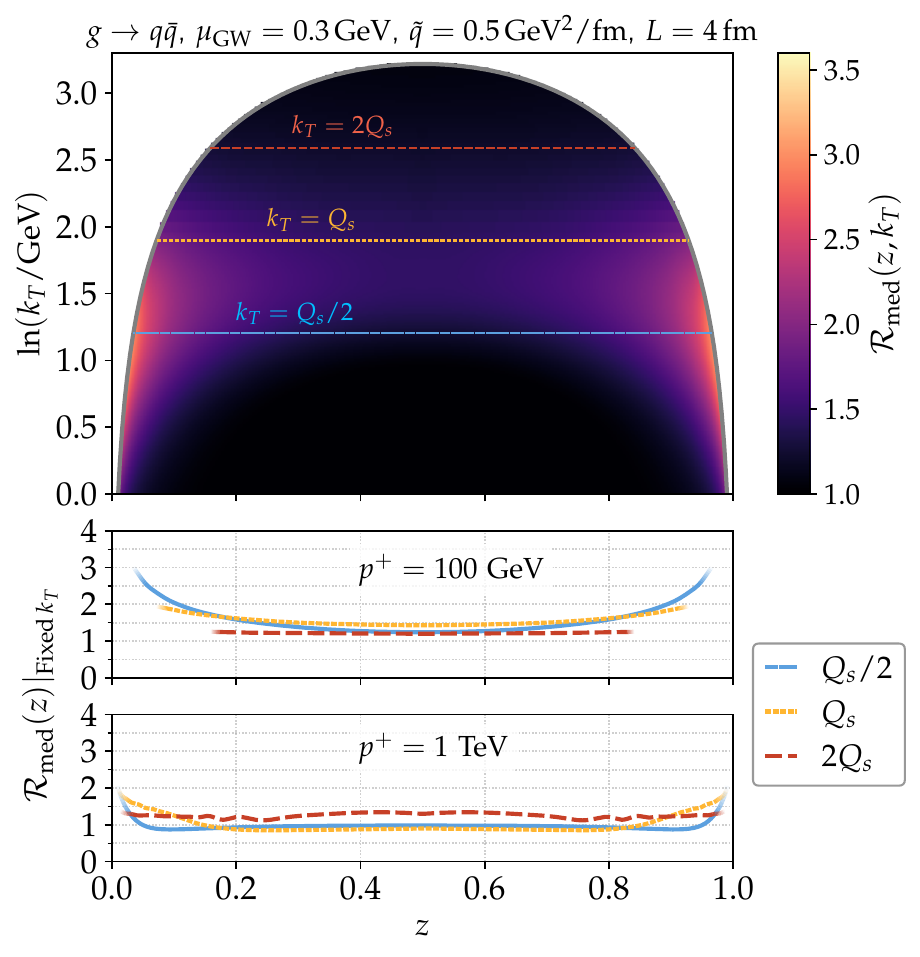}
\caption{Analog of Fig.~\ref{fig:dome} for $g \to gg$ (left) and $g \to q\bar{q}$ (right).}
\label{fig:other_vertices}
\end{figure}

We show the Lund-plane representation of ${\cal R}_{\rm med}(z,k_T)$ for both $g \to gg$ and $g \to q\bar q$ splittings in Fig.~\ref{fig:other_vertices}. The medium parameters and the initial splitter energies are chosen to match those in the main text. We observe that the medium-induced modifications exhibit the same qualitative behavior across all splitting channels. Quantitatively, we find the largest values of ${\cal R}_{\rm med}(z,k_T)$ for $g\to gg$ splittings. We also confirm that the values of ${\cal R}_{\rm med}(z,k_T)$ for $g\to gg$ and $q\to gq$ splittings agree in the $z\to 0$ limit, as explained in~\ref{app:colour-potential}.  Medium modifications to $g\to q\bar q$ splittings remain modest except for highly unbalanced splittings.    
\subsection{Numerical validation}
To rigorously validate the correctness of the algorithm presented in this work, it has been instrumental to benchmark our numerical results against semi-analytic expectations. These analytic baselines are only available when considering two specific approximations: the large-$N_c$ limit and the harmonic-oscillator approximation for the medium scattering potential. The former involves setting $C_A=N_c = 2C_F$ and using Eqs.~\eqref{app:eq:gamma-large-Nc}-\eqref{app:eq:ggg-large-Nc} to determine the color matrices, which is straightforward to do in our code. The latter requires replacing Eq.~\eqref{app:eq:potential} by  
\begin{equation}
\label{eq:vHO-coord}
n\sigma({\bm x}) = \frac{1}{4} \tilde q {\bm x}^2 \, .
\end{equation}
Adapting the algorithm to this concrete form of the potential is less trivial. The main complication arises because the momentum-space representation of Eq.~\eqref{eq:vHO-coord} involves a Laplacian operator acting on a Dirac delta function, given by
\begin{equation}
    \int_{\bm{x}} \bm{x}^2 f(\bm{x}) e^{-i\bm{p}\cdot\bm{x}} = \nabla^2_{\bm{p}}f(\bm{p}).
\end{equation}
Consequently, the momentum-space convolutions in Eqs.~\eqref{eq:schro-A} and \eqref{eq:schro-B} reduce to pure partial derivatives. This violates a fundamental assumption of our algorithm: the continuity of the potential. 

To resolve this issue, we construct a continuous surrogate of the Laplacian operator using a Gaussian kernel, exploiting the identity
\begin{equation}
    \nabla_{\bm{p}}^2 f(\bm{p}) 
= \lim_{\varepsilon \to 0^+}\int \frac{1}{2\pi\varepsilon^4}\left(\frac{\bm{p}'^2}{2\varepsilon^2}-1\right)e^{-\frac{\bm{p}'^2}{2\varepsilon^2}}\,f(\bm{p}-\bm{p}')\, .
\end{equation}
In practice, we run the code for a series of values of $\varepsilon$ and obtain the $\varepsilon \to 0$ limit via numerical extrapolation.
Alternatively, one can  evaluate the necessary derivatives analytically to simplify the expressions prior to numerical computation. This approach requires a dedicated algorithm to solve the resulting equations and we have verified that both methods yield identical results.

\begin{figure}
\includegraphics[width=0.32\textwidth]{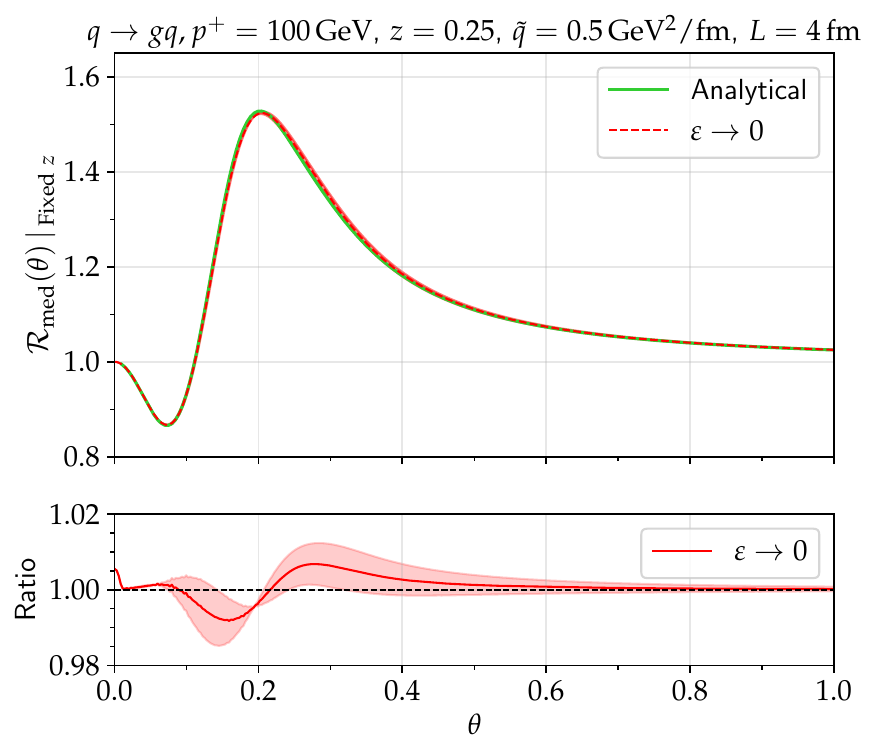}
\includegraphics[width=0.32\textwidth]{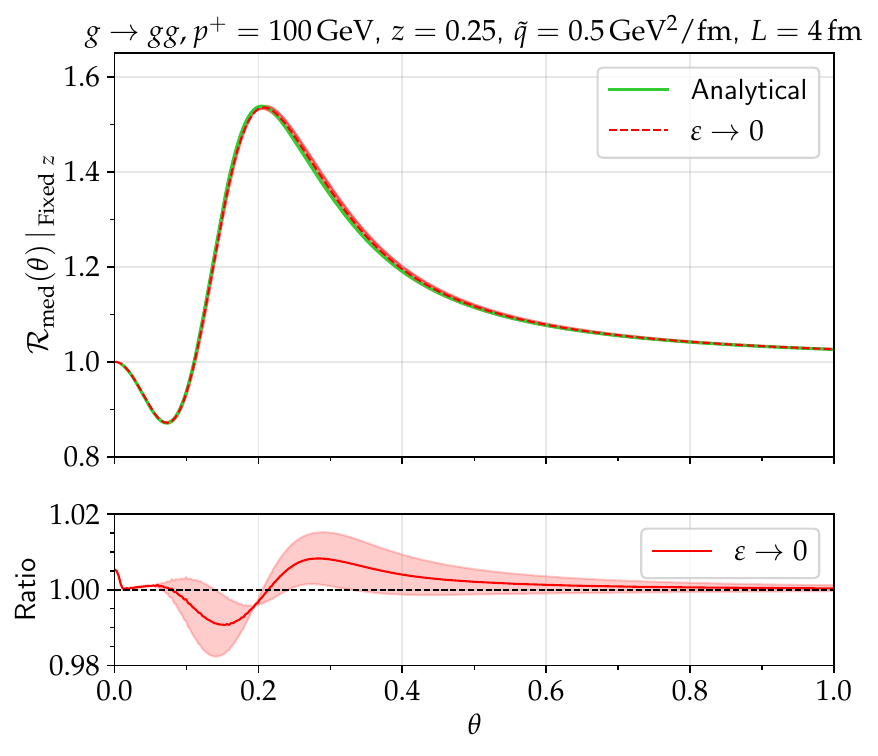}
\includegraphics[width=0.32\textwidth]{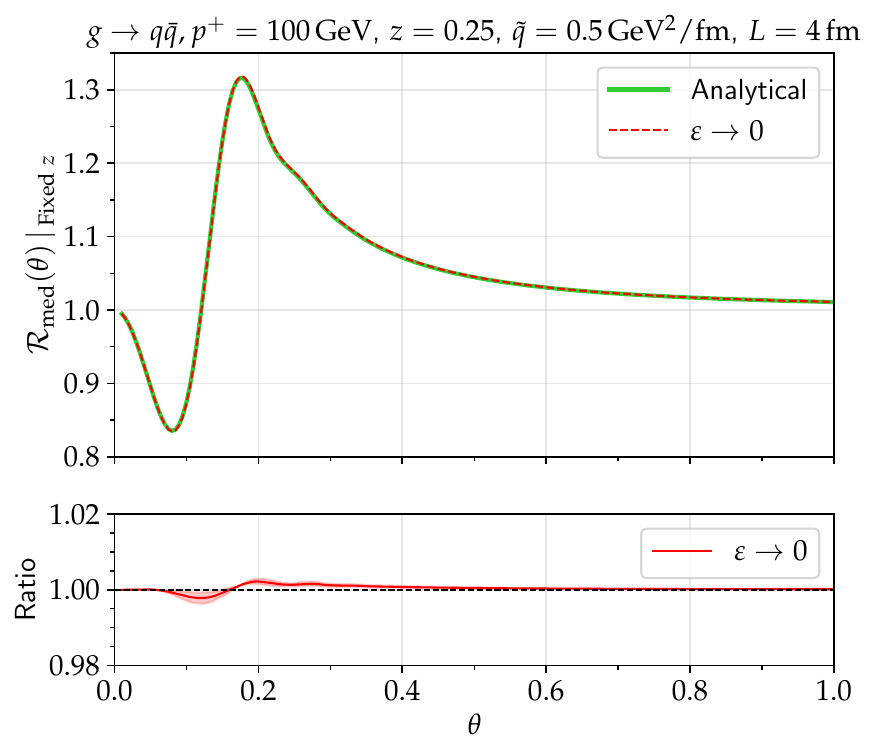}
\caption{Comparison of ${\cal R}_{\rm med}$ between our numerical result and the analytic baseline (using large-$N_c$ and harmonic oscillator approximations) for the three QCD splitting functions and $z = 0.25$. The lower panel shows the ratio between the two results. The band represents the uncertainty due to the numerical $\varepsilon\to 0$ extrapolation.}
\label{fig:comparison-w-largeNc}
\end{figure}

We present a comparison between the $\varepsilon \to 0$ extrapolation and the exact analytic result for ${\cal R}_{\rm med}(z=0.25,\theta=k_T/\omega)$ in Fig.~\ref{fig:comparison-w-largeNc}. The analytic benchmarks are taken from Refs.~\cite{Attems:2022ubu,Andres:2026qrt}. In all channels, the numerical results agree well with the analytic expectation, with deviations consistently bounding below the $1-2\%$ level. We have carried out analogous tests for other medium and kinematic parameters finding robust agreement with the corresponding analytic calculations.

\begin{figure}
\includegraphics[width=0.32\textwidth]{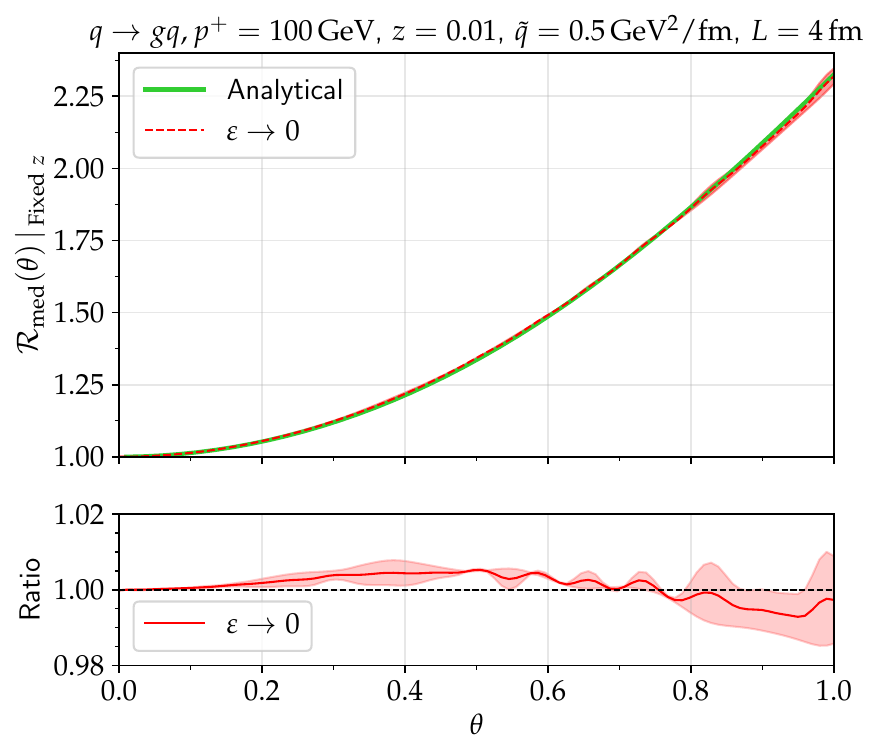}
\includegraphics[width=0.32\textwidth]{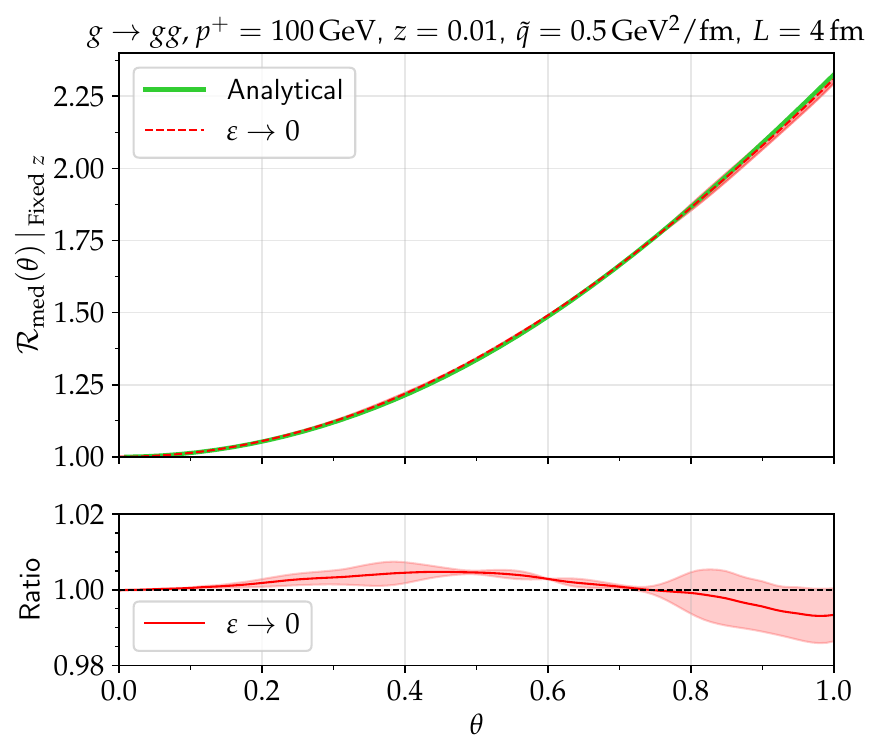}
\includegraphics[width=0.32\textwidth]{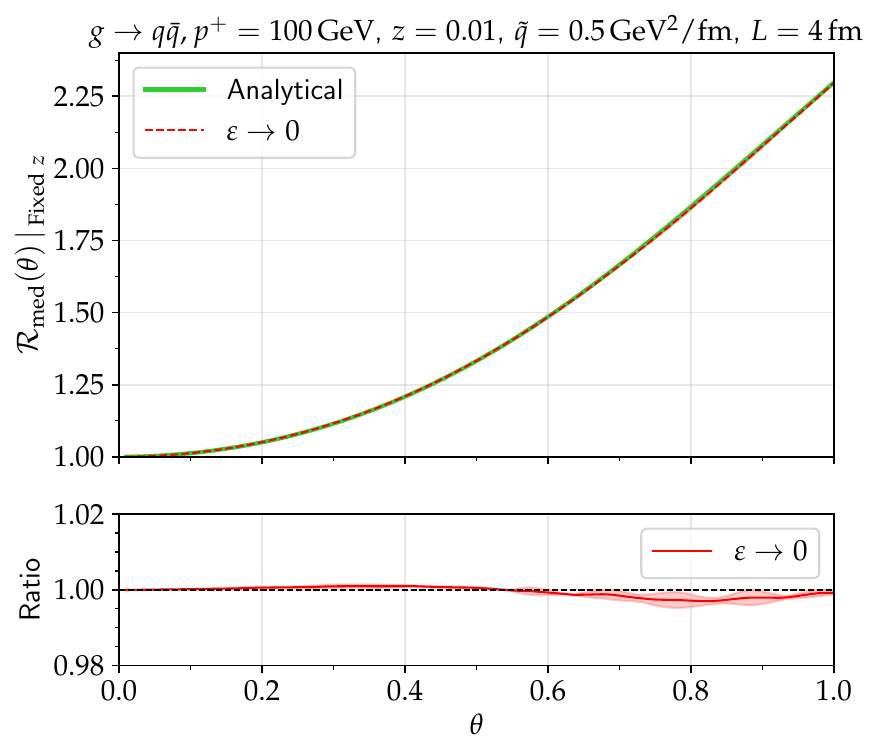}
\caption{Analogue of Fig.~\ref{fig:comparison-w-largeNc} in the soft limit, where we fix $z=0.01$.}
\label{fig:comparison-w-largeNc-and-z0}
\end{figure}

Lastly, in Fig.~\ref{fig:comparison-w-largeNc-and-z0} we perform a similar comparison but for a smaller value of $z=0.01$ and use the soft limit approximation in the analytic results~\cite{Baier:1994bd,Baier:1996kr,Baier:1996sk,Zakharov:1996fv,Zakharov:1997uu,Blaizot:2012fh}, given its widespread phenomenological use. In this limit, numerical simulations and analytic results also agree at the percent level. 

\end{document}